\documentclass[
reprint,
 amsmath,amssymb,
 aps,
prd,
floatfix,
]{revtex4-2}

\usepackage[english]{babel}
\usepackage[utf8x]{inputenc}
\usepackage[dvipsnames]{xcolor}
\usepackage{graphicx}

\usepackage{gensymb}
\usepackage{amssymb}
\usepackage{mathtools}
\usepackage{xspace}
\usepackage{upgreek}
\usepackage{hyperref}
\usepackage{layout}
\usepackage{color}
\usepackage[dvipsnames]{xcolor}
\usepackage{lipsum}
\usepackage{amsmath}
\usepackage{fullpage}

\usepackage{eurosym}
\usepackage{ragged2e}
\usepackage[font=footnotesize, justification=justified, singlelinecheck=false, format=plain, indention=0pt, margin=0pt]{caption}
\DeclareCaptionJustification{justified}{\justifying}
\usepackage{subcaption}
\usepackage{graphicx}
\usepackage{amssymb}
\usepackage{appendix}
\usepackage{placeins}
\usepackage[dvipsnames]{xcolor}

\usepackage[columnwise]{lineno}
\usepackage{comment}

\newcommand{\po}{$^{210}$Po\xspace}
\newcommand{\bi}{$^{210}$Bi\xspace}

\newcommand{\pb}{$^{210}$Pb\xspace}
\newcommand{\be}{$^{7}$Be\xspace}

\newcommand{\pep}{$pep$\xspace}
\newcommand{\pp}{$pp$\xspace}
\newcommand{\cpd}{cpd/100\,t\xspace}

\begin{document}
%\linenumbers
\title{Constraints on non-standard neutrino interactions\\ from Borexino extended data-set}
\author{V.~Antonelli\textsuperscript{1}}
\author{D.~Basilico\textsuperscript{1}}
\author{G.~Bellini\textsuperscript{1}}
\author{J.~Benziger\textsuperscript{2}}
\author{R.~Biondi\textsuperscript{3,a}}
\author{B.~Caccianiga\textsuperscript{1}}
\author{F.~Calaprice\textsuperscript{4}}
\author{A.~Caminata\textsuperscript{5}}
\author{A.~Chepurnov\textsuperscript{6}}
\author{D.~D'Angelo\textsuperscript{1}}
\author{A.~Derbin\textsuperscript{7}}
\author{A.~Di Giacinto\textsuperscript{3}}
\author{V.~Di Marcello\textsuperscript{3}}
\author{X.F.~Ding\textsuperscript{4,b}}
\author{A.~Di Ludovico\textsuperscript{3,4}} 
\author{L.~Di Noto\textsuperscript{5}}
\author{I.~Drachnev\textsuperscript{7}}
\author{D.~Franco\textsuperscript{8}}
\author{C.~Galbiati\textsuperscript{4,9}}
\author{C.~Ghiano\textsuperscript{3}}
\author{M.~Giammarchi\textsuperscript{1}}
\author{A.~Goretti\textsuperscript{4}}
\author{M.~Gromov\textsuperscript{6,22}}
\author{D.~Guffanti\textsuperscript{10,c}}
\author{Aldo~Ianni\textsuperscript{3}}
\author{Andrea~Ianni\textsuperscript{4}}
\author{A.~Jany\textsuperscript{11}}
\author{V.~Kobychev\textsuperscript{12}}
\author{G.~Korga\textsuperscript{13,14}}
\author{S.~Kumaran\textsuperscript{15,16,d}}
\author{M.~Laubenstein\textsuperscript{3}}
\author{E.~Litvinovich\textsuperscript{17,18}}
\author{P.~Lombardi\textsuperscript{1}}
\author{I.~Lomskaya\textsuperscript{7}}
\author{L.~Ludhova\textsuperscript{15,16,e,10}}
\author{I.~Machulin\textsuperscript{17,18}}
\author{J.~Martyn\textsuperscript{10}}
\author{E.~Meroni\textsuperscript{1}}
\author{L.~Miramonti\textsuperscript{1}}
\author{M.~Misiaszek\textsuperscript{11}}
\author{V.~Muratova\textsuperscript{7}}
\author{L.~Oberauer\textsuperscript{19}}
\author{V.~Orekhov\textsuperscript{10}}
\author{F.~Ortica\textsuperscript{20}}
\author{M.~Pallavicini\textsuperscript{5}}
\author{L.~Pelicci\textsuperscript{1}}
\author{\"O.~Penek\textsuperscript{15,f}}
\author{L.~Pietrofaccia\textsuperscript{3}}
\author{N.~Pilipenko\textsuperscript{7}}
\author{A.~Pocar\textsuperscript{21}}
\author{G.~Raikov\textsuperscript{17}}
\author{M.T.~Ranalli\textsuperscript{3}}
\author{G.~Ranucci\textsuperscript{1}}
\author{A.~Razeto\textsuperscript{3}}
\author{A.~Re\textsuperscript{1}}
\author{N.~Rossi\textsuperscript{3}}
\author{S.~Sch\"onert\textsuperscript{19}}
\author{D.~Semenov\textsuperscript{7}}
\author{G.~Settanta\textsuperscript{15,g}}
\author{M.~Skorokhvatov\textsuperscript{17,18}}
\author{A.~Singhal\textsuperscript{15,16,h}}
\author{O.~Smirnov\textsuperscript{22}}
\author{A.~Sotnikov\textsuperscript{22}}
\author{G.~Sun~\textsuperscript{25}}
\author{R.~Tartaglia\textsuperscript{3}}
\author{G.~Testera\textsuperscript{5}}
\author{M.D.C.~Torri\textsuperscript{1}}
\author{E.~Unzhakov\textsuperscript{7}}
\author{A.~Vishneva\textsuperscript{22}}
\author{R.B.~Vogelaar\textsuperscript{23}}
\author{F.~von~Feilitzsch\textsuperscript{19}}
\author{M.~Wojcik\textsuperscript{11}}
\author{M.~Wurm\textsuperscript{10}}
\author{S.~Zavatarelli\textsuperscript{5}}
\author{X.~Zhou\textsuperscript{25}}
\author{K.~Zuber\textsuperscript{24}}
\author{G.~Zuzel\textsuperscript{11}}
% Affiliation
\affiliation{}
\affiliation{\textsuperscript{1}Dipartimento di Fisica, Universit\`a degli Studi e INFN, 20133 Milano, Italy}
\affiliation{\textsuperscript{2}Chemical Engineering Department, Princeton University, Princeton, NJ 08544, USA}
\affiliation{\textsuperscript{3}INFN Laboratori Nazionali del Gran Sasso, 67010 Assergi (AQ), Italy}
\affiliation{\textsuperscript{4}Physics Department, Princeton University, Princeton, NJ 08544, USA}
\affiliation{\textsuperscript{5}Dipartimento di Fisica, Universit\`a degli Studi e INFN, 16146 Genova, Italy}
\affiliation{\textsuperscript{6}Skobeltsyn Institute of Nuclear Physics, Lomonosov Moscow State University, 119234 Moscow, Russia}
\affiliation{\textsuperscript{7}St. Petersburg Nuclear Physics Institute NRC Kurchatov Institute, 188350 Gatchina, Russia}
\affiliation{\textsuperscript{8}APC, Universit\'e de Paris Cit\'e, CNRS, Astroparticule et Cosmologie, Paris F-75013, France}
\affiliation{\textsuperscript{9}Gran Sasso Science Institute, 67100 L'Aquila, Italy}
\affiliation{\textsuperscript{10}Institute of Physics and Excellence Cluster PRISMA+, Johannes Gutenberg-Universit\"at Mainz, 55099 Mainz, Germany}
\affiliation{\textsuperscript{11}M.~Smoluchowski Institute of Physics, Jagiellonian University, 30348 Krakow, Poland}
\affiliation{\textsuperscript{12}Institute for Nuclear Research of NASU, 03028 Kyiv, Ukraine}
\affiliation{\textsuperscript{13}Department of Physics, Royal Holloway University of London, Egham, Surrey,TW20 0EX, UK}
\affiliation{\textsuperscript{14}Institute of Nuclear Research (Atomki), Debrecen, Hungary}
\affiliation{\textsuperscript{15}Institut f\"ur Kernphysik, Forschungszentrum J\"ulich, 52425 J\"ulich, Germany}
\affiliation{\textsuperscript{16}III. Physikalisches Institut B, RWTH Aachen University, 52062 Aachen, Germany}
\affiliation{\textsuperscript{17}National Research Centre Kurchatov Institute, 123182 Moscow, Russia}
\affiliation{\textsuperscript{18}National Research Nuclear University MEPhI (Moscow Engineering Physics Institute), 115409 Moscow, Russia}
\affiliation{\textsuperscript{19}Physik-Department, Technische Universit\"at  M\"unchen, 85748 Garching, Germany}
\affiliation{\textsuperscript{20}Dipartimento di Chimica, Biologia e Biotecnologie, Universit\`a degli Studi e INFN, 06123 Perugia, Italy}
\affiliation{\textsuperscript{21}Amherst Center for Fundamental Interactions and Physics Department, University of Massachusetts, Amherst, MA 01003, USA}
\affiliation{\textsuperscript{22}Joint Institute for Nuclear Research, 141980 Dubna, Russia}
\affiliation{\textsuperscript{23}Physics Department, Virginia Polytechnic Institute and State University, Blacksburg, VA 24061, USA}
\affiliation{\textsuperscript{24}TU Dresden, Dresden, Germany}
\affiliation{\textsuperscript{25}School of Physics and Technology, Wuhan University, Wuhan, China}

% Present Positions
\affiliation{\textsuperscript{a}Present address: Gran Sasso Science Institute (GSSI), L'Aquila, Italy}
\affiliation{\textsuperscript{b}Present address: IHEP Institute of High Energy Physics, 100049 Beijing, China}
% \affiliation{\textsuperscript{c}Present address: INFN Laboratori Nazionali del Gran Sasso, 67010 Assergi (AQ), Italy}
\affiliation{\textsuperscript{c}Present address: Dipartimento di Fisica, Università degli Studi e INFN Milano-Bicocca, 20126 Milano, Italy}
\affiliation{\textsuperscript{d}Present address: Department of Physics and Astronomy, University of California, Irvine, California, USA}
\affiliation{\textsuperscript{e}Present address: GSI Helmholtzzentrum für Schwerionenforschung GmbH, 64291 Darmstadt, Germany}
\affiliation{\textsuperscript{f}Present address: Boston University, Department of Physics, 02215 Boston, MA, USA}
\affiliation{\textsuperscript{g}Present address: Istituto Superiore per la Protezione e la Ricerca Ambientale, 00144 Roma, Italy}
\affiliation{\textsuperscript{h}Present address: WSS-Forschungszentrum catalaix, RWTH Aachen University, Worringerweg 2, 52074 Aachen}

\collaboration{The BOREXINO collaboration -- email: \texttt{spokesperson-borex@lngs.infn.it} }

\date{\today}

\clearpage
\begin{abstract}
Neutrino non-standard interactions (NSI) constitute an active research field, as they are closely related to potential new physics associated with dark matter searches and exotic interactions arising from fundamental symmetry violations.
The Borexino's unprecedented sensitivity to solar neutrinos, derived from its low background and precision spectral measurements, enables stringent constraints on potential deviations from the standard three-flavor neutrino oscillation paradigm. This work presents an update on the analysis of flavor-diagonal NSI using the full Borexino Phase-III data set, extending the study previously reported in \emph{Constraints on flavor-diagonal non-standard neutrino interactions from Borexino Phase-II} (JHEP 2020, 38). The updated analysis incorporates the extended temporal and statistical coverage of Phase-III. The results indicate improved sensitivity to the diagonal NSI parameters, with constraints exceeding those obtained in Phase-II. Furthermore, a more general analysis that includes all possible off-diagonal NSI terms is presented for the first time, providing a comprehensive exploration of the NSI parameter space associated with the flavors of the incoming and outgoing neutrinos. This work once again underlines the Borexino's critical role in probing new physics scenarios and reinforces its legacy in neutrino research. Detailed comparisons with Phase-II results are discussed, along with implications for theoretical models of NSI. 
\end{abstract}

%\keywords{keywords{machine learning, mlti-layer perceptron, artificial neural network, pulse shape discrimination, solar neutrinos, scintillation detectors}

\maketitle

\section*{Introduction}

Non-standard neutrino interactions (NSI) refer to hypothetical interactions between neutrinos and other particles that deviate from the predictions of the Standard Model (SM) of particle physics. Although the SM has been incredibly successful in describing the fundamental particles and their interactions, it does not account for all observed phenomena, such as gravity, dark matter, and dark energy. Besides those fundamental questions, the neutrino plays a special role as a unique portal on the physics beyond SM. 

In particular, NSI go exactly in this direction: investigating them could help us understand the nature of neutrino masses and mixing, provide insights into the origins of matter-antimatter asymmetry, and explore connections to dark matter and other exotic particles related to the possible violation of the axial/vector interaction. 

The Borexino experiment has made significant contributions to science with its discoveries and precise measurements of the nuclear processes that power the Sun and stars~\cite{bib:bell}. With its exceptional low background in the energy range of approximately 150 keV to 15 MeV, Borexino was the only experiment able to detect all the components of solar neutrino spectrum, and, therefore, it greatly contributed to advance our understanding of neutrino oscillations, by measuring simultaneously the survival probability in vacuum and in matter.

In previous studies~\cite{bib:bxnsi}, exploiting data from the detector Phase-II, we analyzed flavor-diagonal NSI, paving the way for understanding how NSI could modify neutrino-electron elastic scattering and the survival probability of solar neutrinos.

Starting from the previous result, we now extend our analysis to a more comprehensive data set including Phase-III, aiming to refine and improve our constraints on NSI. This larger data set offers higher statistics, allowing us to explore more in detail the potential existence of NSI and their implications for neutrino interactions.

In Sec.~\ref{sec:det} we review the main features of the Borexino detector and its most important results in neutrino physics. Furthermore, we review the basic analysis strategy and the characteristics of the extended dataset for the NSI analysis upgrade. In Sec.~\ref{sec:teo} we discuss the theoretical framework of NSI, highlighting the difference of the parameter choice between the old and the new analysis, consisting mainly in the inclusion of the off-diagonal terms. In Sec.~\ref{sec:ana} we discuss the analysis strategy, showing the new Monte Carlo based approach. Finally in Sec.~\ref{sec:res} we report the results of the new analysis, comparing and contrasting it with the old publication and discussing the implications of the results in a wider context including results from other experiments.

This document is followed by a series of appendices, reporting the details of the formalism and calculations used in the main analysis. In order not to deviate the reader from the main content, all those technicalities are reported at the end and can be checked by the most curious readers.

\section{The Borexino detector}
\label{sec:det}

Borexino was mainly detecting neutrinos via the scattering off electrons given by the reaction $\nu_x+e^- \rightarrow \nu_x + e^-$, where $\nu_x$ represents neutrinos of all types interacting with neutral and / or charged current, depending on the flavor according to SM. Borexino was reconstructing the energy, position, and other features of each event in real time.  

The detector was located in Hall C of the Laboratori Nazionali Gran Sasso (LNGS) of the Italian Institute of Nuclear Physics (INFN) \cite{bib:lngs}. The detector had been taking data from mid-2007 to the end of 2021, and is currently under decommissioning.
The detector is made of spherical concentric layers of increasing radiopurity (see \cite{bib:tech} for details): the innermost core, called the Inner Vessel (IV), consists of about 280 tons of liquid scintillator (\emph{pseudocumene} mixed with 1.5 g/l of PPO as scintillating solute) contained inside an ultra-pure nylon vessel with a thickness of 125 $\upmu$m and a radius of 4.25 m. A Stainless Steel Sphere (SSS), filled with the remaining $1000$ m$^3$ of buffer liquid (\emph{pseudocumene} mixed with \emph{dimethylphthalate} quencher) is instrumented with 2212 PMTs to detect scintillation light inside the IV. Ultimately, the SSS is placed inside a water tank (WT) with a volume of 2400 m$^3$, which functions as a Cerenkov veto system, equipped with 200 PMTs. Using results from the study of internal residual contaminations and from the 2010 calibration, it was found that the detector is capable of determining the event position with an accuracy of $\sim 10$ cm (at 1 MeV)  and the event energy with a resolution following approximately the relation $\sigma(E) / E \simeq 5\%/\sqrt{E/[MeV]}$. 

The Borexino data set is divided into three different phases: Phase-I, from May 2007 to May 2010, ended with a calibration campaign, in which the first measurement of the \be solar neutrino interaction rate  \cite{bib:be7-1, bib:be7-2, bib:be7-3} and the first evidence of the \pep solar neutrinos \cite{bib:pep} were achieved; Phase-II, from December 2011 to May 2016, started after an intense purification campaign with unprecedented reduction of the scintillator radioactive contaminants, in which the first spectroscopic observation of the \pp neutrinos with 10\% precision was published~\cite{bib:pp}, and was later updated in the comprehensive analysis of all \pp chain  neutrino fluxes \cite{bib:global, bib:nusol, bib:b8}; finally, Phase-III, from July 2016 to October 2021, started after a thermal stabilization program (consisting of detector insulation and temperature control aimed at stopping the scintillator convective motions) in which the first detection of the CNO neutrinos \cite{bib:cno} and its subsequent improvements \cite{bib:PRLcno, bib:CID} were achieved.

The most important solar neutrino results in terms of interaction rate and corresponding fluxes are summarized in Tab.~\ref{tab:solars}.
Due to its unprecedented radio-purity, Borexino has also set a lot of limits on rare processes, such as potential electron decay
\cite{bib:elec}, NSI \cite{bib:bxnsi}, low energy neutrinos correlated with astrophysical events
\cite{bib:astro-nu}, neutrino magnetic moment
\cite{bib:magnet}, sterile neutrino \cite{bib:sterile},
and performed other studies, such as \emph{e.g.} the determination of very stringent limits for possible violations of Pauli's principle in nuclear systems \cite{Borexino-Pauli} and geo-neutrino detection \cite{bib:geonu}, which gives essential contributions to the testing and discrimination of possible geophysical models for the Earth.
\begin{table*}[!t]
%\nolinenumbers
\centering
\begin{tabular}{ccc}
\hline\hline
Species & Rate [\cpd] & Flux [cm$^{-2}$ s$^{-1}$ ] \\ \hline
{\it pp}     & $(134 \pm 10)^{+6}_{-10}$  & $(6.1 \pm 0.5)_{-0.5}^{+0.3} \times 10^{10}$ \\
$^7$Be       & $(48.3 \pm 1.1)^{+0.4}_{-0.7}$ & $(4.99 \pm 0.11)_{-0.08}^{+0.06} \times 10^9$ \\
{\it pep} (HZ) & $(2.7 \pm 0.4)^{+0.1}_{-0.2}$ & $(1.3 \pm 0.3)_{-0.1}^{+0.1} \times 10^8$ \\ 
$^8$B ($>3$ MeV) & $0.223_{-0.022}^{+0.021}$ & $5.68_{-0.44}^{+0.42} \times 10^6$ \\
{\it hep}    & $<0.002$ (90\% CL) & $<1.8 \times 10^5$ (90\% CL) \\
CNO          & $6.7_{-0.8}^{+1.2}$ & $6.7_{-0.8}^{+1.2} \times 10^8$ \\
\hline\hline
\end{tabular}
\caption{Solar neutrino interaction rates in Borexino and extrapolated solar neutrino fluxes for the different components of the \pp chain and CNO cycle. Rates are reported in \cpd, while fluxes are reported in \texorpdfstring{$\text{cm}^{-2}\,\text{s}^{-1}$}{cm$^{-2}$s$^{-1}$}. N.B.: HZ stands for high metallicity assumption. N.B.: \cpd is the standard Borexino rate unit and stands for counts per day per 100 tonnes.}
\label{tab:solars}
%\linenumbers
\end{table*}

\subsection{Data selection}

For the most important analyses in Borexino, the fundamental event selection is based on the following criteria: internal only trigger (no muon veto coincidence), event time greater than 2\,ms from a preceding muon (to remove cosmogenics), a single cluster in the acquisition window, and the position typically reconstructed in $r < 2.8$\,m and $-1.8 < z < 2.2 $\, m, where $r$ is the distance from the detector center and $z$ is the vertical position. These cuts guarantee that the selected event is a neutrino-like candidate, i.e. an event occurred in the innermost part of the IV ($\lesssim$100 tonnes around the center) and far enough from the external background that comes from the SSS and from the IV structures. 

After applying the selection criteria listed above, the typical Borexino spectrum shows a prominent $^{210}$Po $\alpha$ peak at about 500 keV, which falls within the $^7$Be energy window; see \emph{e.g.}~\cite{bib:be7-2}.  
At the beginning of Phase-I, the \po activity was of the order of $10^4$ \cpd\footnote{Standard Borexino rate unit (\cpd stands for \emph{counts per day per 100 tonne.})}. At the beginning of Phase-II, more than 4 years later, the activity decreased by one order of magnitude to $\sim10^3$ \cpd, 
a bit more than expected because the water extraction campaign reintroduced a small amount of \po.
Finally, in Phase-III, after more than 4 years from the Water Extraction, the scintillator convective motions were drastically reduced thanks to a thermal insulation campaign. This significantly reduced \po activity by another order of magnitude, reaching $\sim 20$ \cpd. This allowed Borexino to reach the \po condition required for the CNO measurement \cite{bib:cno} via the $^{210}$Bi independent constraint from the secular equilibrium, in the jargon \bi-\po  \emph{link}.
\begin{table*}[t!]
%\nolinenumbers
\begin{center}
\begin{tabular}{cccc}
\hline \hline 
 & Phase-I  & Phase-II & Phase-III \\ \hline
$^{210}$Bi & $\sim 42$ cpd/100t & $\sim 18$ cpd/100t & $\sim 11$ cpd/100t \\
$^{210}$Po & $\sim 700$ cpd/100t & $\sim 250$ cpd/100t & $< 20$ cpd/100t \\
$^{85}$Kr & $\sim 35$ cpd/100t &  $< 6.8$ cpd/100t  (95\% CL) &   $<11$ cpd/100t (95\% CL) \\
$^{238}$U & $(5.3\pm0.5)\times10^{-18}$ g/g &  $<9.4 \times 10^{-20}$ (95\% CL) & $(6.5\pm 2.2) \times 10^{-20}$ \\
$^{232}$Th & $(3.8\pm0.8)\times10^{-18}$ g/g &  $<5.7 \times 10^{-19}$ (95\% CL) & $<2.8 \times 10^{-19}$ (95\% CL)\\ \hline \hline
\end{tabular}
\end{center}
\caption{Comparison between the rates of the major backgrounds for solar neutrino analysis in the fiducial volume, before (Phase-I) and after the purification campaign (Phase-II), and after the thermal stabilization campaign in Phase-III (N.B.: \po reported here is an average over the entire phase).}
\label{tab:majorbkg}
%\linenumbers
\end{table*} 

From Phase-II and Phase-III the content of \bi was reduced by a 60\% due to the thermal stability program for the CNO detection. The higher value estimate in Phase-II is largely due to top/bottom asymmetry followed after the last purification campaign. At the beginning of Phase-II a very clean scintillator portion was left on the detector top hemisphere by the latest purification cycles. Thanks to convective motions, activated by thermal instabilities before the insulation program, the full scintillator bulk has been extensively mixed leading to a uniform distribution of \pb, the \bi parent nuclide, as described in~\cite{bib:cno, bib:PRLcno}.

The $^{85}$Kr, strongly reduced (more than a factor 5) after the intensive purification campaign in between Phase-I and Phase-II, has always showed a residual rate limit, quantified thanks to the $\beta-\gamma$ delayed coincidence decay channel, whose branching ratio is 0.43\%. This limit will be reviewed and discussed in Sec.~\ref{sec:ana} as it may play an important role in the NSI interaction. The rest of contamination (as the cosmogenics, the external background and the uranium and thorium decay chain daughters) are basically constant or negligible; see Sec.~\ref{sec:ana} as well.

\subsection{The extended dataset}

The analysis presented in this document is based on data taken when the radiopurity and thermal stability of the detector was maximal, i.e., between
July 2016 and October 2021 (Phase-III). The last
part of the dataset features an unprecedented thermal
stability and an enlarged volume of strongly reduced
$^{210}$Po contamination (see Table 2),
%(see Fig. 1), 
and therefore provides
an improved $^{210}$Po constraint, useful as will be discussed in Sec.~\ref{sec:ana}. The
total exposure of the analysis presented in this document is about 
1500 days × 70 tons.

\section{Theoretical framework}
\label{sec:teo}

\subsection{Motivations and possibilities for the study 
of Non Standard neutrino Interactions}
The possible existence of NSI, in addition to the usual neutral and charge current electroweak interactions predicted by the SM is under investigation in the literature since many years \cite{first-NSI}. Originally \cite{NSI-Nunokawa} they had been advocated to find an alternative explanation of the solar neutrino data to the usual oscillation mechanism, introducing the possibility of massless neutrino oscillations. 
Throughout the years, the sensitivity to possible NSI has been analyzed for different neutrino sources: atmospheric neutrinos \cite{NSI-atmospheric-Fornengo} 
reactor neutrinos \cite{NSI-reactors, NSI-DAYA-BAY, NSI-T2K} %
and, more recently, long baseline accelerator experiments, LBL \cite{NSI-LBL}, which pose particular attention to the possible influence of NSI on CP violation sensitivity \cite{NSI-LBL-CP, NSI-LBL-CP-2} and neutrino mass ordering discrimination power \cite{NSI-LBL-ordering, JUNO-yellow-book}.

The theoretical interest for NSI remained high over the years 
\cite{NSI-BSM-2009, Large-NSI-2009}, also in connection with dark matter models, that often predict their existence. Different NSI coefficients, parametrizing possible extensions of neutrino interaction with matter, have been analyzed in a series 
of papers, among which recent studies reported in \cite{Concia-first, Concia-mass-hierarchy, Concia-with-Coherent, Concia-quarks-and-electrons,Concia-CP}. 

Oscillation data typically constrain the differences between diagonal neutral current NSI parameters. To derive additional constraints on the single NSI parameters one can use the fact that eventual NSI corrections would influence also the inelastic neutrino scattering and perform a combined analysis of oscillation and scattering data, incorporating in the analysis also the limits coming from medium and high energy electroweak precision data \cite{FASER}. 
Additional constraints, as in the case of \cite{Large-NSI-2009}, can also be obtained from processes involving
four charged fermions and from decays of muons and mesons.

The most stringent limits, which affect the parameters related to NSI involving incoming and outgoing muonic neutrinos (usually denoted as $\varepsilon_{\mu\,\mu}$ coefficient), can be recovered from the data
of CHARM~\cite{bib:charm} at CERN.

However, new models have been introduced in the literature (for instance, \cite{Farzan-2015}) in which the NSI arise by the interaction with light mediators not affected by the limits from high energy neutrino experiments, like CHARM and NuTeV.
The constraints surviving also in these light mediator models have been studied by combining the oscillation data with those derived by the neutrino nucleus scattering experiments, like COHERENT \cite{COHERENT}, using neutrinos from the Spallation Neutron Source (SNS) at Oak Ridge National Laboratory and another experiment with neutrinos from the Dresden-II reactor \cite{Concia-quarks-and-electrons}.

\subsection{Investigating NSI with Borexino data: potentiality and novelties of the analysis} 

\label{analysis-potentiality-and-novelty}
The Borexino experiment offers the opportunity of monitoring the full solar neutrino spectrum, including the measurements of all the pp-chain components and also measuring for the first time the CNO neutrinos.
From the point of view of NSI searches an ideal situation is represented by the high statistic measurement of the
$^7 {\rm Be}$ signal, which has the advantage of being mono-energetic and, therefore, of having a clear expected experimental signature. 

In this work, we will focus on the so-called vector-NSI, a specific form of NSI that introduces terms in the interaction Lagrangian with matter, modifying the conserved currents. Besides, there are also the so-called scalar-NSI, which perturb the mass eigenstates, again due to modifications in the interaction with matter \cite{SNSI}. Generally speaking, the possible vector NSI coefficients can be written as a sum over different contributions corresponding to neutrino interactions with leptons and quarks, respectively. The introduction of NSI can foresee the possibility for all neutrino flavors to interact with ordinary matter leptons via the charged current (CC) channel. The effective interaction Lagrangian density valid at tree level is obtained combining neutral and charged current contributions. The CC Lagrangian can be written using a Fierz reordering in order to obtain a form comparable to the neutral current (NC) one. In the effective framework, the total interaction Lagrangian density at the leading order is modified by the introduction of a NSI depending term:
\begin{equation}
\label{NSI-chiral}
\mathcal{L}_{NSI}=-2\sqrt{2}G_{F}\sum_{\substack{\alpha, \beta \\ l, l', X}} \varepsilon_{\alpha\beta}^{ll'X}\left(\overline{\nu}_{\alpha}\gamma_{\mu}P_{L}\nu_{\beta}\right)\left(\overline{l}\gamma^{\mu}P_{X}l'\right)
\end{equation}
In the above expression, the chiral index $X=L,\,R$ labels the left and right NSI contributions. The indices $\alpha,\,\beta=e,\,\mu,\,\tau$ denote the outgoing and incoming neutrino flavors, while $l$ and $l'$ represent the outgoing and incoming charged leptons or quarks carrying the same electric charge.
Requiring the interaction Lagrangian to be Hermitian imposes a constraint on the $\varepsilon$ coefficients:
\begin{equation}
\varepsilon_{\alpha\beta}^{l\,l'X}=\left(\varepsilon_{\alpha\beta}^{l\,l'X}\right)^{\dagger}.
\end{equation}
The chirality projectors are defined as usual as:
\begin{align}
\label{chirality}
P_{L}=\frac{1}{2}(1-\gamma_{5}),\quad
P_{R}=\frac{1}{2}(1+\gamma_{5})
\end{align}
where $P_{X}$ corresponds to the value of the index $X$ specifying the chirality of the $ll'$ lepton current.

Experiments like Borexino are sensitive only to vector NSI that are diagonal in the flavor of the incoming and outgoing lepton ($l = l' = e$). An analysis based on Borexino results has already been performed \cite{bib:bxnsi}, limited to lepton-diagonal NSI and deriving limits on both the left- (L) and right-handed (R) NSI coefficients ($\varepsilon_{\alpha\alpha}^{L}$ and $\varepsilon_{\alpha\alpha}^{R}$).

The complete NSI coefficients, representing in the Hamiltonian the full effect of NSI corrections, including 
the contributions coming from all the possible fermionic interactions, can be written as the product of a common (flavor and chirality independent) coefficient and two factors, $\chi^X$ and $\psi^l$ depending, respectively, on the chirality and on the charged lepton flavor, that is:
$
\varepsilon_{\alpha\beta}^{l\,l\,X} = 
\varepsilon_{\alpha\beta} \, \psi^l \chi^X \, .
$
Due to the presence of the chiral index in Eq.~(\ref{NSI-chiral}), the NSI corrections can be parametrized in terms of left and right coefficients $\varepsilon_{\alpha\beta}^{l\,l\, L/R}$, which can also be rearranged, giving NSI vector (V) and axial (A) coefficients, defined as:
\begin{equation}
\varepsilon_{\alpha\,\beta}^{l\,l\; V/A} = \frac{1}{2} 
\left[\varepsilon_{\alpha\beta}^{l\,l\, R} \pm \varepsilon_{\alpha\beta}^{l\,l\, L}\right] 
\label{def-vector-axial}
\end{equation}
In this paper the vector and the axial NSI coefficients have been studied and relative limits have been recovered separately for these two extreme cases, focusing on the scenario in which the charged lepton participating in the interaction is the electron ($l = e$) and, therefore, the index $l$ of Eq.~(\ref{def-vector-axial}) will be omitted in the following.

In this work we are updating and extending the NSI studies with Borexino data and the main novelties with respect to the previous analysis \cite{bib:bxnsi} are the use of a larger statistical data set, with the inclusion of Borexino phase III data, and, from a theoretical point of view, the extension to possible NSI off diagonal in the neutrino flavors. 
 
It has already been shown in \cite{bib:bxnsi} that the impact of possible NSI on neutrino production is negligible, while they can be relevant for neutrino propagation (through modifications of the oscillation probability in matter) and, mainly, for the neutrino interaction cross sections in the detector. 
Therefore, in this work, we developed a complete analysis, including all possible sources of NSI related to the incoming and outgoing flavor of the neutrino involved in the interaction, deriving explicitly the modified oscillation probabilities and the modified interaction cross sections. 
In this way, we obtained the first fully comprehensive study of vector NSI with solar neutrino data.

\subsection{Neutrino propagation}
\label{propagation}

In the context of the Hamiltonian formalism, neutrino propagation is described by the vacuum Hamiltonian matrix summed to the contribution given by the matter interaction operator. In order to classify the different possible non standard neutrino interactions we can start from the following expression for the Hamiltonian ruling the neutrino propagation in matter and determining also the interactions in the detector:

\begin{equation}
\begin{split}
\label{nsihamiltstandard}
&\mathcal{H} = \mathcal{H}_{Vac} + \mathcal{H}_{matt} =\\
&=U  \frac{1}{2E_{\nu}} \left(\begin{array}{ccc}
0 & 0 & 0 \\
 & \Delta m_{21}^2 & 0 \\
0 & 0 & \Delta m_{31}^2 \\
\end{array} 
\right)U^{\dagger}  +
\mathcal{H}_{matt}\,. 
\end{split}
\end{equation} 

In the previous Eq.~(\ref{nsihamiltstandard}) $\Delta m_{ij}^2$ denotes, as usual, the differences among the squares of the three neutrino mass eigenvalues. The matrix $U$ appearing in Eq.~(\ref{nsihamiltstandard}) 
is the usual Pontecorvo-Maki-Nakagawa-Sakata (PMNS) unitary transformation matrix.

\begin{figure}[htbp]
    %\nolinenumbers
    \centering
    \includegraphics[width=0.9\linewidth]{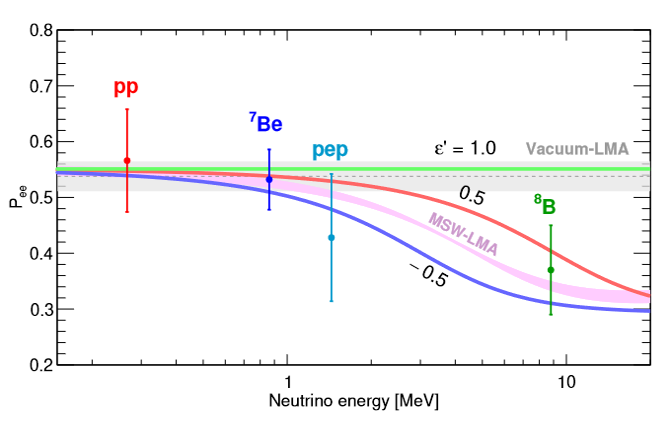}
    \caption{Electron neutrino survival probability $P_{ee}(E)$ as a function of neutrino energy.\cite{bib:bxnsi}}
    %\linenumbers
    \label{fig:nsipee}
\end{figure}

Even in the presence of NSI, only the flavor-dependent part of the matter potential affects neutrino oscillations. Flavor-universal NC-like NSI contributions generate a common phase for all neutrino flavors which can be removed by a redefinition of the neutrino phases. Consequently, the effective NSI contribution relevant for oscillations can be treated as a CC-like matter potential. The non-standard interaction corrections are included in the term $\mathcal{H}_{matt}$, which is a function of the Fermi coupling constant $G_F$ and of the electron density in matter $N_e$. $H_{matt}$ can be parametrized in terms of the different NSI parameters in the following way: 
\begin{equation}
\label{1}
\mathcal{H}_{matt}=
\sqrt{2}G_{F}N_{e}
\left(
  \begin{array}{ccc}
    1+\varepsilon_{ee} & \varepsilon_{e\mu} & \varepsilon_{e\tau} \\
    \varepsilon_{\mu e} & \varepsilon_{\mu\mu} & \varepsilon_{\mu\tau} \\
    \varepsilon_{\tau e} & \varepsilon_{\tau\mu} & \varepsilon_{\tau\tau} \\
  \end{array}
\right)
\end{equation}
where $N_{e}$ is the averaged electron density of the considered material and the coefficients $\varepsilon_{\alpha\beta}$ are the sum of the left and right contributions. The hermiticity of the previous Hamiltonian matrix requires that $\varepsilon_{\alpha\beta}=\varepsilon^{*}_{\beta\alpha}$, as a consequence the diagonal terms must be real ($\varepsilon_{\alpha\alpha}\in \mathbb{R}$), whereas the off-diagonal terms must be the complex conjugates of the corresponding transposed ones:
\begin{equation}
\begin{split}
\label{1a-bis}
&\mathcal{H}_{matt}=\\
&=\sqrt{2}G_{F}N_{e}
\left(
  \begin{array}{ccc}
    1+\varepsilon_{ee} & \varepsilon_{e\mu}e^{i\phi_{e\mu}} & \varepsilon_{e\tau}e^{i\phi_{e\tau}} \\
    \varepsilon_{e\mu}e^{-i\phi_{e \mu}} & \varepsilon_{\mu\mu} & \varepsilon_{\mu\tau}e^{i\phi_{\mu\tau}} \\
    \varepsilon_{e\tau}e^{-i\phi_{e\tau}} & \varepsilon_{\mu\tau}e^{-i\phi_{\mu\tau}} & \varepsilon_{\tau\tau} \\
  \end{array}
\right).
\end{split}
\end{equation}
In order to preserve the hermiticity of the Hamilton operator, in the case of off-diagonal coefficients the left and right phases have to be equal $\phi_{\alpha\beta L}=\phi_{\alpha\beta R}$:
\begin{equation}
\begin{split}
&\varepsilon_{\alpha\alpha}=\varepsilon_{\alpha\alpha L}+\varepsilon_{\alpha\alpha R}\\
&\varepsilon_{\alpha\beta}e^{i\phi_{\alpha\beta}}=\varepsilon_{\alpha\beta L}e^{i\phi_{\alpha\beta L}}+\varepsilon_{\alpha\beta R}e^{i\phi_{\alpha\beta R}}=\\
&\qquad\qquad=\left(\varepsilon_{\alpha\beta L}+\varepsilon_{\alpha\beta R}\right)e^{i\phi_{\alpha\beta}}
\end{split}
\end{equation}
In the numerical analysis we performed, we adopted the simplifying assumption of real NSI and, hence, we put to zero all the 
$\phi_{\alpha\beta}$ coefficients. This corresponds to assuming that the NSI coefficients are identical for neutrinos and antineutrinos and that no additional CP-violating phases are introduced, while the difference in the matter interaction structure between neutrinos and antineutrinos is preserved.
In this work, the considered NSI are diagonal in the flavor of the charged leptons interacting with neutrinos, that is, the interactions involve incoming matter electrons, and the outgoing charged leptons are again electrons.
In the context of the introduced Hamiltonian formalism, the different oscillation probabilities can be computed, as reported in the Appendix \ref{appendixA}. For example, the survival probability of $\nu_{e}$ becomes: 
\begin{equation}
P_{ee}=\sin^4{\theta_{13}}+\frac{1}{2}\cos^4{\theta_{13}}\left(1+\cos{(2\theta_{M})}\cos{(2\theta_{12})}\right),
\end{equation}
$\theta_{M}$ is the matter mixing angle for the Sun, modified by the introduction of the NSI parameters:
\begin{equation}
\begin{split}
\label{thetaM}
\tan{(2\theta_{M})}&=\tan{(2\theta_{12})}\bigg[1-\frac{2\,V_{matt}\cdot\varepsilon_{e\mu}^{V}\cdot\cos{\theta_{23}}}{\Delta m_{21}^{2}\sin{(2\theta_{12})}}\bigg]\\
&\bigg[1-\frac{V_{matt}(1-\varepsilon')}{\Delta m_{12}^{2}\cos{(2\theta_{12})}}\bigg]^{-1},
\end{split}
\end{equation}
where we have introduced the parameter 
\begin{equation}
\label{epsilon}
\varepsilon'=-\varepsilon_{ee}^{V}+\varepsilon_{\mu\mu}^{V}\cos^{2}{\theta_{23}}+\varepsilon_{\tau\tau}^{V}\sin^{2}{\theta_{23}}.
\end{equation}
and $\varepsilon_{\alpha\beta}^{V}=\varepsilon_{\alpha\beta}^{L}+\varepsilon_{\alpha\beta}^{R}$. The modified survival probability as a function of neutrino energy is reported in Fig.~\ref{fig:nsipee}, obtained by introducing NSI parameter $\varepsilon'=\pm0.5$.

\subsection{Neutrino-matter NSI interaction Lagrangian and NSI modified cross section}

After a Fierz reordering, CC-like NSI operators can be cast into an effective NC-like form relevant for neutrino–matter interactions and the interaction Lagrangian including NSI can be written as: 
\begin{equation}
\begin{split}
\mathcal{L}=&-2\sqrt{2}~G_{F} \, \sum_{\alpha}\left(\overline{\nu}_{\alpha}\gamma_{\mu}P_{L}\nu_{\alpha}\right)\\
&\big[\left(g_{\alpha L}+\varepsilon_{\alpha\alpha L}\right)\left(\overline{e}\gamma^{\mu}P_{L}e\right)+\left(g_{\alpha R}+\varepsilon_{\alpha\alpha R}\right)\\
&\left(\overline{e}\gamma^{\mu} P_{R}e\right)\big]
-2\sqrt{2}G_{F}\sum_{\alpha\neq\beta}\left(\overline{\nu}_{\alpha}\gamma_{\mu}P_{L}\nu_{\beta}\right)\\
&\left[\left(\varepsilon_{\alpha\beta L}\right)\left(\overline{e}\gamma^{\mu}P_{L}e\right)+\left(\varepsilon_{\alpha\beta R}\right)\left(\overline{e}\gamma^{\mu}P_{R}e\right)\right].
\end{split}
\end{equation}

The left and right contributions have been split using the L-R coupling constants defined as:
\begin{subequations}
\begin{align}
g_{\alpha L}&=
\begin{cases}
g_{LL}^{\alpha e}+1& \text{for flavor $\alpha=e$}\\
g_{LL}^{\alpha e}& \text{for flavor $\alpha=\mu,\,\tau$}
\end{cases}\\
g_{\alpha R}&=\;\;g^{\alpha e}_{LR}\qquad\quad \text{for every flavor $\alpha$}.
\end{align}
\end{subequations}
The coupling constants are defined as:
\begin{subequations}
\begin{align}
g_{LL}^{\alpha e}=\frac{1}{2}(g_{LV}^{\alpha e}+g_{LA}^{\alpha e}),\quad
g_{LR}^{\alpha e}=\frac{1}{2}(g_{LV}^{\alpha e}-g_{LA}^{\alpha e}).
\end{align}
\end{subequations}
At the leading order, the vectorial and axial coupling can be simply computed:
\begin{subequations}
\begin{align}
g_{LV}^{\alpha e}=-\frac{1}{2}+2\sin^2{\theta_{W}},\quad
g_{LA}^{\alpha e}=-\frac{1}{2}
\end{align}
\end{subequations}
where $\theta_{W}$ is the Weinberg angle. As a final result, it is possible to obtain the explicit form of the L-R coupling constants:
\begin{subequations}
\begin{align}
g_{LL}^{\alpha e}=-\frac{1}{2}+\sin^2{\theta_{W}},\quad
g_{LR}^{\alpha e}=\sin^2{\theta_{W}}.
\end{align}
\end{subequations}
Starting from the previous results, the NSI-modified $\nu$--$e$ interaction cross section, as a function of the recoiled electron kinetic energy $T$, for a neutrino beam of energy $E_\nu$, can be written as:
\begin{equation}
\label{crosssec1}
\frac{d\sigma_{\alpha\beta}}{dt}=
\frac{1}{16\pi}
\frac{\left|\mathcal{M}_{\alpha\beta}\right|^{2}}
{\left(s-m_{e}^{2}\right)^2} \, ,
\end{equation}
where an incoming neutrino of flavor $\alpha$ scatters into a neutrino of flavor $\beta$ through its interaction with an electron. Here $\mathcal{M}_{\alpha\beta}$ denotes the corresponding matrix element. In the relativistic limit with the electron at rest, the Mandelstam variables are given by
$s=m_e^2+2m_eE_\nu$ and $t=-2m_eT$.
Using $dt/dT = -2m_e$, one obtains:
\begin{equation}
\frac{d\sigma_{\alpha\beta}}{dT}=
\frac{\left|\mathcal{M}_{\alpha\beta}\right|^{2}}{32\pi m_e^2 E_\nu^2} \, .
\end{equation}
The cross section for the interaction of a neutrino of flavor $\alpha$ with an electron can finally be written in the form:
\begin{equation}
\label{cross1}
\begin{split}
&\frac{d\sigma_{\alpha}(E,\,T)}{dT}
=\frac{2}{\pi}G_{F}^2m_{e}\bigg[\big[\left(g_{\alpha L}+\varepsilon_{\alpha\alpha L}\right)^2\\
&+\sum_{\beta\neq\alpha}\left(\varepsilon_{\alpha\beta L}\right)^2\big]+\big[\left(g_{\alpha R}+\varepsilon_{\alpha\alpha R}\right)^2\\
&+\sum_{\beta\neq\alpha}\left(\varepsilon_{\alpha\beta R}\right)^2\big]\left(1-\frac{T}{E}\right)^2\\
&-\bigg((g_{\alpha L}+\varepsilon_{\alpha\alpha L})(g_{\alpha R}+\varepsilon_{\alpha\alpha R})\\
&+\sum_{\beta\neq\alpha}(\varepsilon_{\alpha\beta L}\varepsilon_{\alpha\beta R})\bigg)\left(\frac{m_{e}T}{E^2}\right)\bigg].
\end{split}
\end{equation}
where the contributions given by the different outgoing neutrino flavors have been summed up.\\
It is important to note that the cross section involves only the magnitude of the NSI parameters. Hence, the choice of real or complex off-diagonal NSI parameters does not affect the final form of the differential cross section.\\
The cross section obtained reduces to the one used in the previous work \cite{bib:bxnsi} in the case of null off-diagonal NSI terms.
Using the previous results, it is possible to obtain the scattered electron spectrum in the form:
\begin{equation}
\begin{split}
\label{spectrum1}
\frac{dR}{dT}=&N_{e}\Phi_{\nu}\int dE\frac{d\lambda_{\nu}}{dE}\bigg[\frac{d\sigma_{e}}{dT}P_{ee}(E)\\
+&\frac{d\sigma_{\mu}}{dT}P_{e\mu}(E)+\frac{d\sigma_{\tau}}{dT}P_{e\tau}(E)\bigg].
\end{split}
\end{equation}
where the oscillation probability $P_{\alpha\beta}$ has been introduced.\\
Including the three-flavors MSW analysis framework it is possible to obtain the expected recoiled electron spectrum as a function of the incoming neutrino flux energy.

\begin{figure}[htbp]
        \centering
        %\nolinenumbers
        \includegraphics[width=0.9\linewidth]{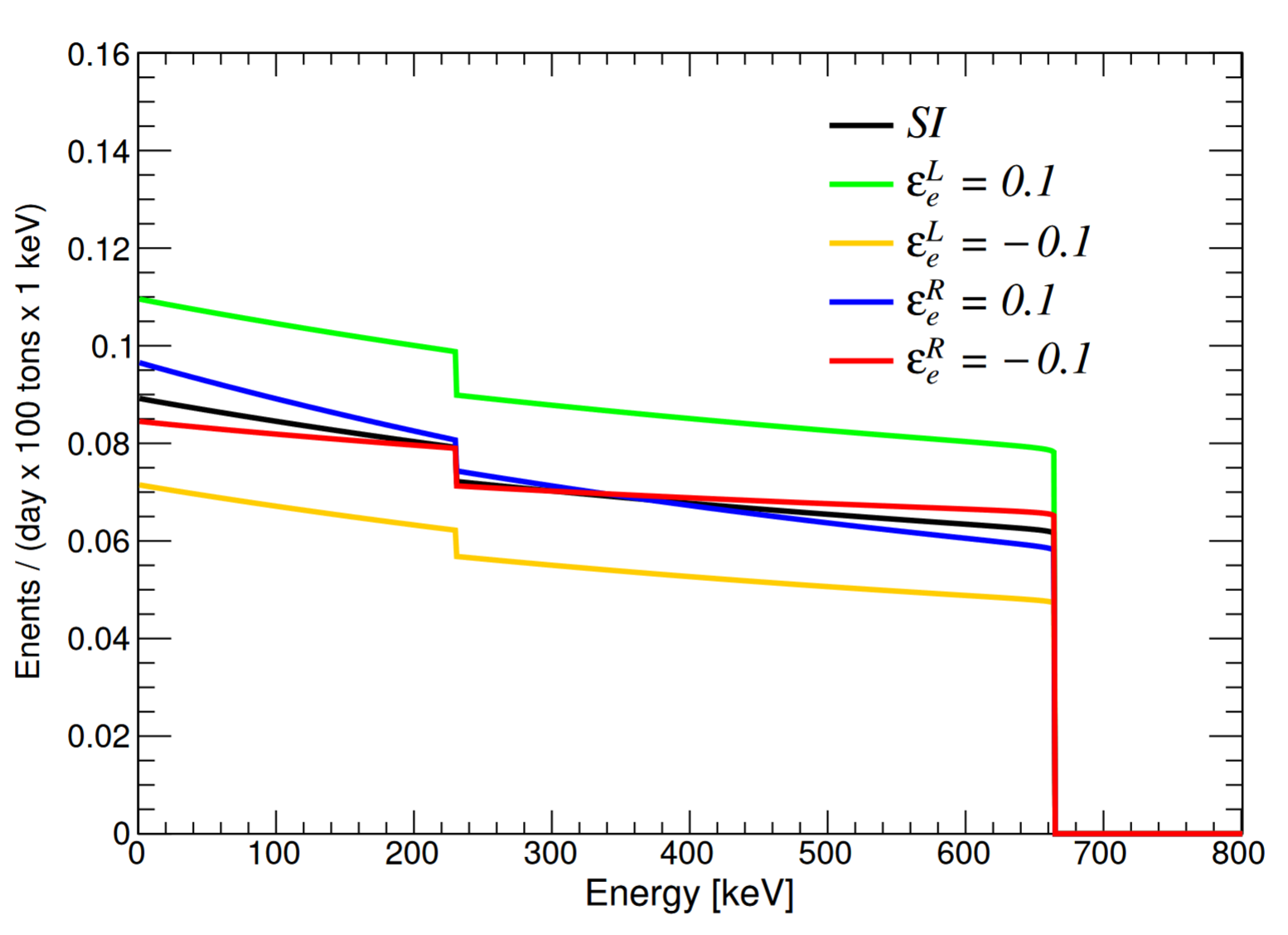}
        \caption{Distortion of the electron recoil spectrum for $^7\rm{Be}$ solar neutrino due to non-zero values of $\varepsilon_{e}^{L/R}$ without detector resolution.}
        %\linenumbers
        \label{figxsec}
\end{figure}

\section{Data Analysis}
\label{sec:ana}

This chapter discusses in detail how the NSI analysis, based on Borexino Phase-II and Phase-III data, is performed. Basically, the electron recoil spectra induced by each solar neutrino are modified according to NSI and eventually fitted to the data according to the standard fitting algorithms used in all the Borexino's previous analyses. The most crucial part, and indeed the novelty of the present analysis consisting of the production of the NSI probability density functions(PDFs), is discussed in the following subsections.

\subsection{NSI modified survival probability and cross sections}

As we have largely discussed in the introductory chapters, 
NSI has the potential to exert influence on various aspects of neutrino physics, including production, propagation, and detection. Since Borexino mainly focus on the elastic scattering process between electrons and neutrinos, we mainly focus on the detection aspects.
The dynamics of neutrino propagation in a medium is predominantly governed by vectorial combinations of NSI parameters, represented as $\varepsilon_{\alpha\beta}^{V}=\varepsilon_{\alpha\beta}^{L}+\varepsilon_{\alpha\beta}^{R}$. The graphical representation in Fig.~\ref{fig:nsipee} elucidates a simplified depiction of the impact of NSI-induced matter effects on the electron neutrino survival probability $P_{ee}(E)$. Specifically, the parameter $\varepsilon'$ (Eq.~(\ref{epsilon})) plays a crucial role in modulating this effect\footnote{For a complete discussion on the oscillation probability and the definition of the parameter $\varepsilon^{\prime}$ see Appendix \ref{appendixA}.}. As $\varepsilon^{\prime}$ approaches unity, the matter-induced potential diminishes, causing $P_{ee}(E)$ to converge towards the scenario where all solar neutrinos are oscillating within the vacuum regime. In contrast, in the case where $\varepsilon^{\prime}<0$, the survival probability $P_{ee}(E)$ experiences a reduction as a function of energy.
    
As shown in Fig. \ref{figxsec}, NSI alter the kinetic energy spectrum of recoil electrons. In particular, the figure shows the distortion of the electron recoil spectrum induced by the $^7\rm{Be}$ solar neutrino due to non-zero values of $\varepsilon_{ee}^{L/R}$. In this example the effect of the finite energy resolution of the detector is not included \cite{bib:bxnsi}.

The NSI modified cross section with $\varepsilon^{L/R}$ parameterization can be expressed using Eq.~(\ref{cross1}) and can be used to obtain the electron recoil spectrum as a function of the incoming neutrino flavor; see Eq. (\ref{spectrum1}). Here we can preliminarily observe that $\varepsilon^{L}$ mainly influences the normalization of the recoiled electron spectrum, while $\varepsilon^{R}$ mainly influences the slope. Fig.~\ref{figxsec} shows a vivid explanation for our conclusion. In the following Subsection, we are going to discuss the next step, that is, the further change of the electron recoil spectrum after including the detector response and resolution.

\subsection{Detector response and resolution through \emph{reweighting}}

Starting from the theoretical electron recoil spectra, the MC simulates the detector response and resolution on an event-by-event basis. The Borexino spectrum, as a function of the energy estimator \texttt{nhits} is typically fitted to MC PDFs produced with a customized \textsc{Geant4} simulation, known in Borexino jargon as \texttt{G4Bx}\cite{G4bx}.
Making the same procedure for all possible distortions of every spectra by NSI (following the change of each possible $\varepsilon$' values) is going to be a cumbersome enterprise with many costs in terms of computing power and analysis time. Instead, a faster, but not less accurate, procedure consisting of PDFs \emph{reweighting} has been adopted. 

\begin{figure}[htbp]
    \centering
    %\nolinenumbers
    \includegraphics[width=1\linewidth]{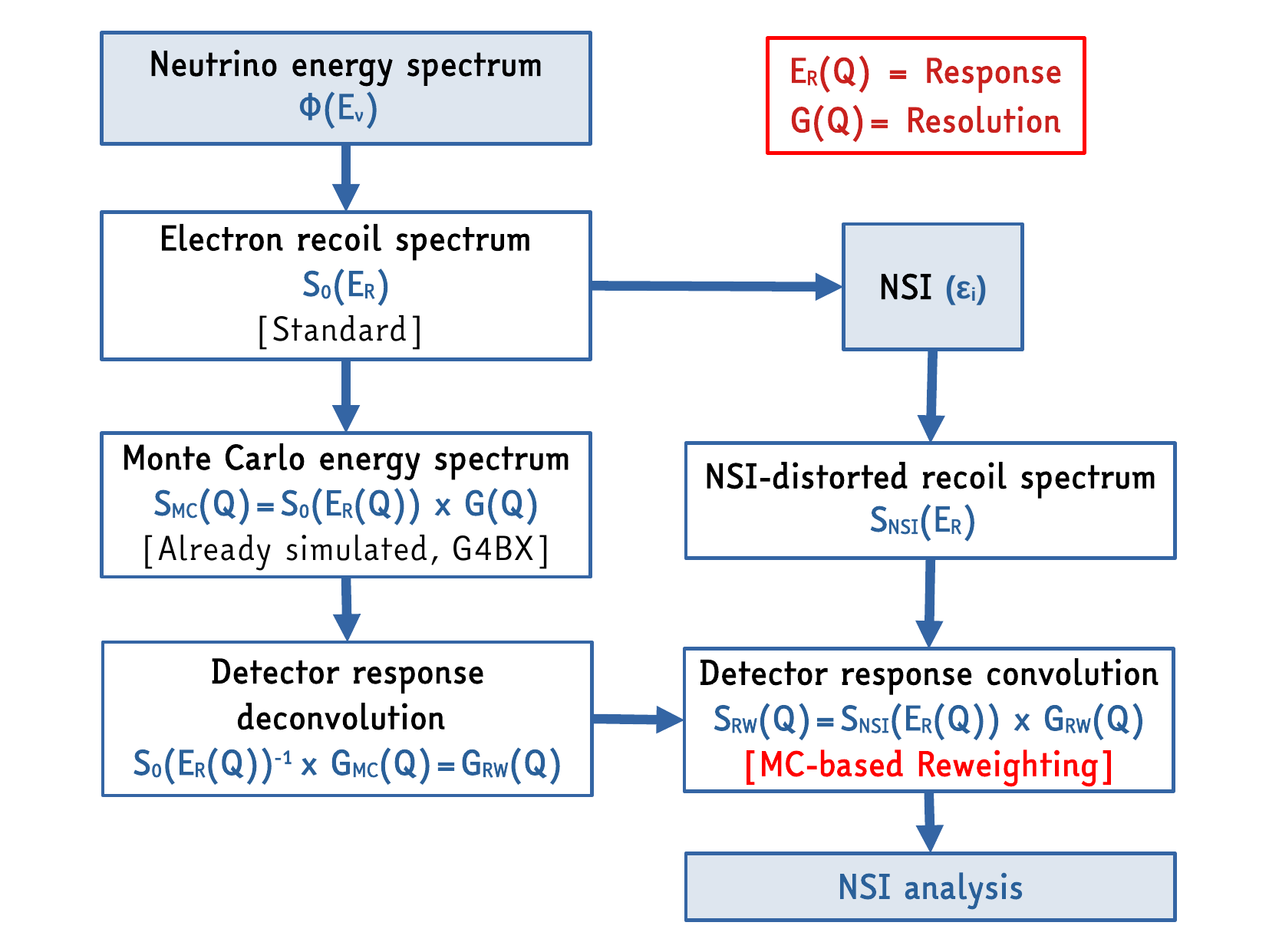}
    \caption{Flow chart of the \emph{reweighting} process. Starting from the PDFs of electron recoil spectra simulated with \texttt{G4Bx}, the detector response is deconvoluted comparing the final energy deposition to the Monte Carlo truth. The obtained response function is later convoluted to the NSI-distorted energy spectra. The final spectra are then used in the NSI spectral fit.}
    \label{fig:blueprint}
    %\linenumbers
\end{figure}

A crucial step in this procedure involves manipulating the PDFs derived from \texttt{G4Bx} to obtain information about the detector response and resolution. Let $S_0(E_R)$ be the electron recoil ($E_R$) spectrum of a given solar neutrino species without any NSI deformation and without detector response and resolution. The MC PDFs, $S_{MC}(Q)$, as a function of the observed energy estimator $Q$ (\texttt{nhits}, in our case), of each species can be understood as a convolution with detector response, namely
\begin{equation}
    S_{MC}(E_R) = S_0(E_R(Q)) \otimes G(Q),
\end{equation}
where $E_R(Q)$ and $G(Q)$ represent in symbols the detector response and resolution, respectively.  The \emph{reweighting} can be determined by deconvolving the detector response $G_{RW}(Q)$, \emph{i.e.} in symbols:
\begin{equation} \label{eq:deco}
    G_{RW}(Q) = S_0(E_R(Q))^{-1} \otimes S_{MC}(Q) \, ,
\end{equation}
where the negative exponents represent the operation of reverting the convolution procedure. In general, the deconvolution is not mathematically defined by a direct analytic formula; rather, it depends on the type of deconvolution adopted. In our specific case, since we know the original simulated recoil energy (MC truth), we can deconvolve the response function bin-by-bin from the two-dimensional histogram of \texttt{nhits} (Q) versus $E_R$.
 
The final electron recoil spectrum containing the NSI information, including  the detector response  and resolution, is obtained by performing a MC convolution of the NSI-deformed theoretical spectrum $S_{NSI}(E_R)$ with $G_{RW}(Q)$ extracted in Eq.~(\ref{eq:deco}), namely,
 \begin{equation}
    S_{RW}(Q) = S_{NSI}(E_R(Q)) \otimes G_{RW}(Q) \, .
\end{equation}
These reweighted PDFs are then utilized to calculate the expected spectra, which are essential for fitting the Borexino data in search of any possible non-zero NSI effect, as a function of the parameters of interest.
Fig.~\ref{fig:blueprint} illustrates the flow chart of the \emph{reweighting} procedure.

Being a novelty of the NSI analysis, the \emph{reweighting} procedure has required a modification of the standard fitting algorithm used in the standard Borexino analysis. For this reason, the new fitting code (Borexino Fitter, or Fitter in short) has required a long period of validation, comparing and contrasting the new results with the previously published ones, as will be described below.

\subsection{Validation of the fitting code}

The NSI analysis of the Borexino Phase-II data has  previously been published \cite{bib:bxnsi}, providing a valuable reference for our fitting program. Except for the modifications required  by the NSI analysis, the Fitter is based on the old Borexino's traditional fitting algorithm, including all the features used in the previous publications by the Collaboration for all the analyses concerning solar neutrinos. To perform NSI analysis by integrating data from two phases, we have developed a fitting program utilizing \textsc{CUDA} \cite{cuda}, a parallel computing platform and application programming interface (API) model developed by \textsc{NVIDIA}\cite{nvidia}.

The Fitter validation steps will be discussed in the following Subsections.

\subsubsection{Fitting Phase-II and Phase-III data without NSI}
\begin{figure}[htbp]
    \centering
    %\nolinenumbers
    \includegraphics[width=0.9\linewidth]{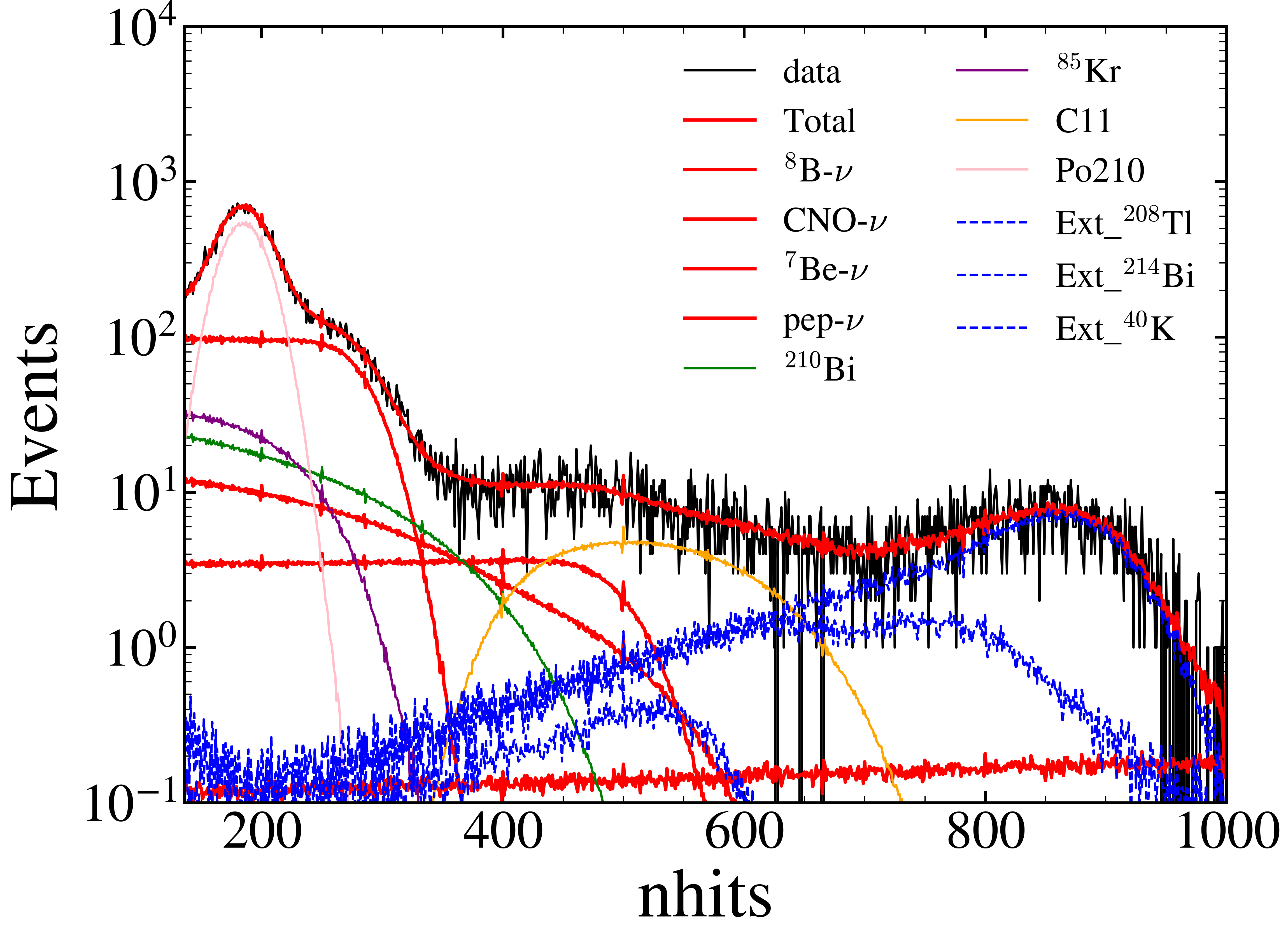}
    \caption{Example of the fitting result of Borexino energy spectrum. }
    \label{fig:P3spectrum}
    %\linenumbers
    \end{figure}

The initial validation in our analysis program involved fitting the spectra while including NSI effects. The results of this fitting process have been previously documented in \cite{bib:global} and \cite{bib:cno}. Building upon this foundation, we conducted a similar fitting analysis without accounting for NSI while using our newly developed fitting program. Subsequently, we compared our results, specifically focusing on the event rates for each spectral component, with the previously published findings. Through a meticulous bin-by-bin cross-check, we found that our fitting results align closely with the published results at a high precision level of $10^{-9}$. The energy spectrum obtained is depicted in Fig. \ref{fig:P3spectrum}.

\subsubsection{Cross check with previous NSI analysis}
\begin{figure*}[htbp]
%\nolinenumbers
        \centering
        \begin{subfigure}{0.4\linewidth}
        \centering
        \includegraphics[width=\textwidth]{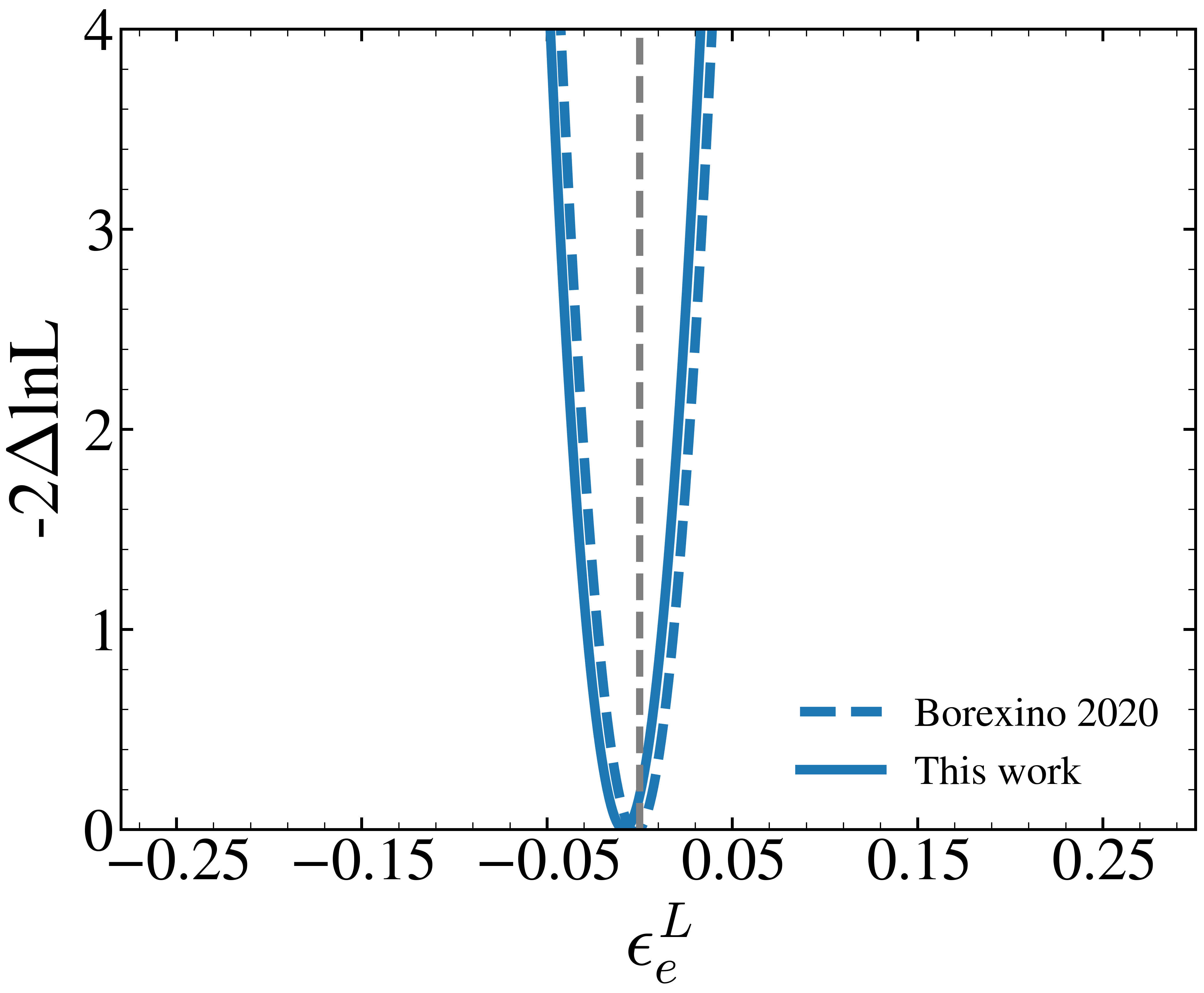}
        \end{subfigure}
        \begin{subfigure}{0.4\linewidth}
        \centering
        \includegraphics[width=\textwidth]{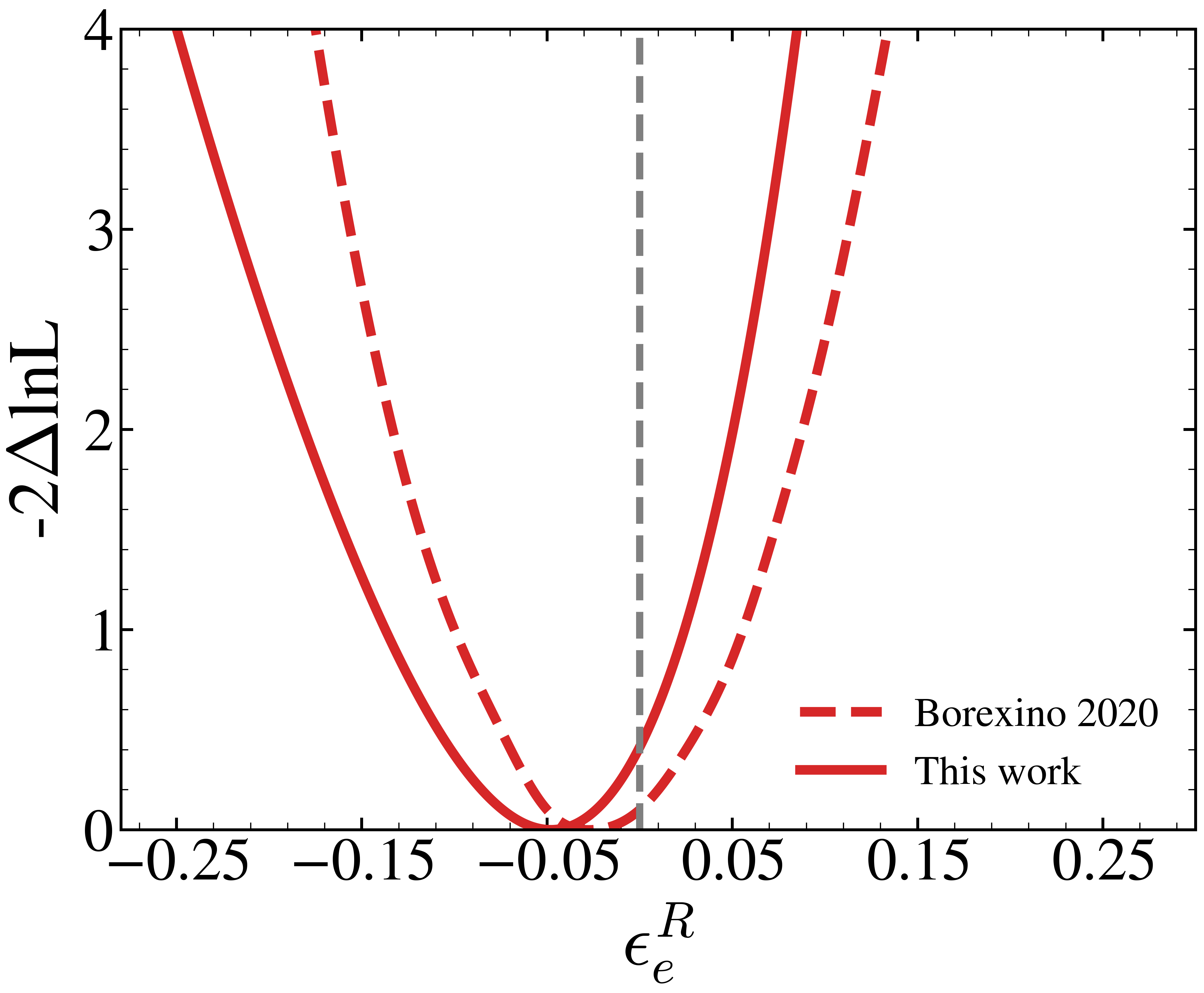}
        \end{subfigure}
        \caption{Comparison of 1-dimensional profiling $\varepsilon_{ee}^{eeL/R}$, the solid lines represent our result and the dashed lines show the published constrain~\cite{bib:bxnsi}.}
        \label{fig:eeLRprofile_p2}
%\linenumbers
\end{figure*}

Our aim is first to replicate the same analysis strategy as outlined in \cite{bib:bxnsi} and to compare the constrained results. Ref.~\cite{bib:bxnsi} reports the NSI constraints based on Borexino Phase-II data, specifically focusing on the flavor-diagonal terms of NSI parameters $\varepsilon_{ee}^{L/R}$ and $\varepsilon_{\tau\tau}^{L/R}$. One minor deviation is our choice to utilize Monte Carlo (MC) simulation to model the detector response, in contrast to the analytical functions used in the previous NSI analysis. Fig.~\ref{fig:eeLRprofile_p2} shows the $-2\Delta \ln L$ profiles of $\varepsilon_{ee}^{L/R}$ from this work and the previously published NSI constraint. Based on the dataset,  we produce similar constraints for $\varepsilon_{ee}^{L/R}$. The results are basically compatible at a level better than $1\sigma$, as it can be inferred from the $-2\Delta \ln L$ profiles,  except for a small deviation originating from the different detector modeling, already observed in many past analyses as a source of systematics.

\subsection{Choice of fitting parameters}
Generally speaking, distinct sets of fitting parameters are required for each solar neutrino flux, background component rate, and NSI coefficients when fitting data from both Phase-II and Phase-III. When amalgamating the datasets from these two phases to constrain NSI effects, certain fitting parameters can be shared between the two datasets. It is assumed that the neutrino fluxes and NSI coefficients do not change over time, thereby allowing for the sharing of these specific fitting parameters across both Phase-II and Phase-III datasets. In contrast, parameters related to background rates and coefficients describing backgrounds are treated separately, as they are not consistent across the two phases. There are three kinds of fitting a total of 44 parameters in our analysis:
\begin{itemize}
\item {\bf NSI parameters:}\\ 
A range of 12 NSI coefficients $\varepsilon_{\alpha\beta}^{X} (\alpha\beta=ee, e\mu, e\tau, \mu\mu, \mu\tau, \tau\tau; X=L/R)$ are employed as fitting parameters according to our analysis strategy outlined in Sec.~\ref{LRparameterization} and Sec.~\ref{VAparameterization}. Depending on the specific analysis procedure, the NSI parameters can be adjusted accordingly.
\item {\bf Solar neutrino fluxes:}\\
We considered 5 solar neutrino fluxes: $pp, ^{8}\rm{B}, \rm{CNO}, ^{7}\rm{Be}, pep$.
While fitting Phase-II data, $pp, pep, ^8\rm{B}, ^7\rm{Be}, and\ \rm{CNO}$ solar neutrino are considered in our analysis, instead, for fitting Phase-III data, $pp$ solar neutrino is excluded as \cite{bib:cno} suggests. In Phase-III indeed, the loss of PMT and the consequent loss of resolution does not allow us to fit the very low energy region.
Since we have mentioned that the neutrino fluxes are assumed to remain constant for Borexino Phase-II and Phase-III, we can use the same solar neutrino flux parameters while fitting different datasets.
\item {\bf Backgrounds:}\\
A total of 25 background coefficients have been employed.\\
For Phase-II data: $^{14}\rm{C}$, $^{11}\rm{C}$, $^{10}\rm{C}$, $^{6}\rm{He}$,$^{210}\rm{Po}$, $^{85}\rm{Kr}$, $^{210}\rm{Bi}$, $^{208}\rm{Tl},$ $^{214}\rm{Bi}$, $^{40}\rm{K}$, and moreover we need an additional fitting parameter to describe $^{14}\rm{C}$ pile up and two additional for TFC subtracted $^{210}\rm{Po}, ^{11}\rm{C}$. Therefore there are totally 13 background parameters for Phase-II data. Phase-III data share the same strategy, but $^{14}\rm{C}$ and $pp$ solar neutrino are not considered since low energy parts including $^{14}\rm{C}$ and $pp$ are not used for Phase-III spectra fitting; an additional parameter is added for the migration of $^{210}\rm{Po}$, as a result of which there are 12 background parameters for Phase-III.
On an event-by-event basis, solar neutrino-induced events are indistinguishable from background events caused by decays of $\beta$ and $\gamma$ in the scintillator. Those backgrounds are divided into three types: internal backgrounds, external backgrounds, and cosmogenic backgrounds. The main source of internal backgrounds are gammas from isotopes contained in the liquid scintillators, such as $^{85}\text{Kr}$ and $^{210}\text{Pb}$ sub-chain. As for external backgrounds, there are two sources: gammas from isotopes in the surrounding environment such as $^{208}\text{Tl}$, $^{214}\text{Bi}$, $^{40}\text{K}$,
and cosmogenic backgrounds such as $^{11}\text{C}$, $^{10}\text{C}$, $^{6}\text{He}$. 
\item {\bf 2 coefficients} used for the $^{85}\rm{Kr}$ penalty for Phase-II and Phase-III. These two fitting parameters were free to fluctuate between $[0,1]$ as the detection efficiency of $^{85}\rm{Kr}$ if applying $^{85}\rm{Kr}$ penalty. When the penalty is neglected, these two parameters are fixed at zero.
\end{itemize}
       
\subsection{Independent penalties on backgrounds and solar fluxes} \label{penalty}

The NSI analysis has been improved considering independent pieces of information on some specific backgrounds, available thanks to the longstanding strategies developed by the Collaboration in previous important analysis. In particular, in the RoI of the NSI fit, two important contaminants can play a special role: $^{210}$Bi and $^{85}$Kr. Both the energy spectra of these $\beta$-emitters are, with different degrees, degenerate in shape with the electron recoil spectra of solar neutrinos, therefore an independent prior of them can constrain the resulting degeneracy in the spectral fit.
Independent constraints on those specific contaminants are reviewed and described in the following Subsections.  

Finally, the independent constraint on solar neutrino fluxes from the Standard Solar Model (SSM), crucial for gaining the NSI sensitivity,  especially for contributions resulting only in an overall normalization effect, is described in the last Subsection.
    
\subsubsection{$^{210}$Bi penalty}
The quantity of ${ }^{210} \mathrm{Bi}$ present in the scintillator is determined by the minimum value of the ${ }^{210} \mathrm{Po}$ rate $R\left({ }^{210} \mathrm{Po}_{\text {min }}\right)$ within the so-called \emph{low polonium field} (LPoF), as expressed by the relationship:
$$
R\left(^{210} \mathrm{Po}_{\text{min}}\right)=R\left(^{210}\mathrm{Bi}\right)+R\left(^{210}\mathrm{Po}^{V}\right),
$$
where the rate of $^{210} \mathrm{Bi}$ is equivalent to $^{210} \mathrm{Po}^{S}$($^{210} \mathrm{Po}$ from the scintillator) according to secular equilibrium. Since $^{210} \mathrm{Po}^{V}$($^{210} \mathrm{Po}$ from the vessel) is consistently positive, the $^{210}\mathrm{Po}_{\text{min}}$ value provides an upper bound for the $^{210}\mathrm{Bi}$ rate.

The distribution of ${ }^{210} \mathrm{Po}$ within LPoF is spatially nonuniform, exhibiting a distinct minimum without a significant plateau in its vicinity. This characteristic establishes a reliable upper limit for the ${ }^{210} \mathrm{Bi}$ rate; however, it does not confirm that ${ }^{210} \mathrm{Po}^{V}$ is precisely zero. Only a spatially extended minimum in the ${ }^{210} \mathrm{Po}$ rate would have allowed the determination of the ${ }^{210} \mathrm{Bi}$ rate.

\begin{table*}[!t]
%\nolinenumbers
    \centering
    \scriptsize
    \begin{tabular}{clccccccc}
    \hline
    \hline $\text { Name }$ & k & b & m $\text { (tons) }$ & T $\text { (days) }$ & $\varepsilon_{\mathrm{BR}}$ & $\varepsilon_{\operatorname{det}}$ & $\sigma$ & Rate(cpd/100t)  \\
    \hline $\text{ Maximum exposure }$ & 8 & 2.03 & 156 & 1780 & 0.43 \% & 20.5 \% & 15 \% & 1.24\\
    \text { Strong prompt cut } & 6 & 1.69 & 156 & 1915 & 0.43 \% & 13.7 \% & 15 \% & 1.72\\
    \text { PeriodAll (2011 Dec - 2016 May) } & 3 & 0.47 & 156 & 1299 & 0.43 \% & 13.6 \% & 15 \% & 2.06\\
    \text { Phase3Final(2016 June - 2021 Oct) } & 6 & 0.54 & 156 & 1570 & 0.43 \% & 10.8 \% & 15 \% & 4.74\\
    Updated\ Phase2\ only & 7 &  0.69 &  156.17 &  1307.3 &  0.43\% &  19.72\% &  15 \% & 6.70\\ 
    Updated\ Phase3Final\ only &6 &  0.54 &  156.17 &  1570.7 &  0.43\% &  8.03`\% &  15 \% & 6.56\\ 
    \hline
    \hline
    \end{tabular}
    \caption{The values of parameters in $^{85}$Kr penalty for different datasets.}
    \label{tab:krpenalty}
%\linenumbers
\end{table*}

The minimum rate of $^{210} \mathrm{Po}$ was deduced from the distribution of ${ }^{210} \mathrm{Po}$ within the LPoF field utilizing 2D and 3D fitting procedures following two mutually compatible methodologies. The stable spatial position of the minimum throughout the analysis period, gradually shifting by less than $20 \mathrm{~cm}$ per month, indicates that the detector operates under a fluid-dynamical quasi-steady-state condition and that the minimum in the ${ }^{210} \mathrm{Po}$ rate is not merely a statistical fluctuation.
In fact, this result was the major breakthrough that emerged from the thermal insulation campaign. 

Both procedures consistently yield $R(^{210}\rm{Po_{min}})=11.5\pm1.0$ cpd/100t.\cite{bib:cno}
\subsubsection{$^{85}$Kr penalty}
The quantity of \%$^{85}\rm{Kr}$-coincidence events generally depends on the specific cuts applied and the exposure duration. Due to the limited number of $^{85}\rm{Kr}$ candidates, the Poisson likelihood method is employed for its penalty assessment. The initial expression of the Poisson likelihood is given by:
\begin{equation}
    -\log (\mathcal{L})_1 = -k \cdot \log (\lambda + b) + \log \Gamma(k + 1) + (\lambda + b)
\end{equation}

where \( k \) represents the observed number of events, \( b \) is the expected number of background events, and \( \lambda \) denotes the expected number of signal events, which can be expressed as:
\begin{equation}
    \lambda = R \cdot m \cdot T \cdot \varepsilon_{\mathrm{BR}} \cdot \varepsilon_{\mathrm{det}}
\end{equation}

Here, \( R \) is the rate of ${ }^{85} \mathrm{Kr}$, \( m \cdot T \) is the exposure, \( \varepsilon_{\mathrm{BR}} \) is the branching ratio and \( \varepsilon_{\mathrm{det}} \) is the detection efficiency.

To account for systematic uncertainties in the selection efficiency of coincidences, we introduce a nuisance parameter \( \varepsilon \) and incorporate a pull term. The modified likelihood function is defined as:
\begin{equation}
\begin{split}
    -\log (\mathcal{L})_1 &= [-k \cdot \ln (\lambda \cdot \varepsilon + b) + \ln \Gamma(k + 1) \\
    &+ (\lambda \cdot \varepsilon + b)] + \frac{(\varepsilon - 1)^2}{\sigma_\varepsilon^2}
\end{split}
\end{equation}
where \( \sigma_\varepsilon \) represents the precision of the efficiency parameter \( \varepsilon \).

Tab.~\ref{tab:krpenalty} presents the values of the $^{85}$Kr penalty parameters for various Borexino datasets.

\subsubsection{Standard Solar Model penalty}

Since some of the NSI deformations are degenerate with the absolute solar flux normalization, a reasonable independent constraint on solar fluxes is necessary. For this reason, we add penalty factors to constrain neutrino fluxes to SSM predictions\cite{bib:SSM}, whose uncertainties for many fluxes are larger than the experimental errors available today.
In particular, we adopted the B16(GS98)-HZ SSM (high metallicity model) to constrain solar fluxes according to Tab. \ref{tab:ssm_flux}. The penalty term reads:
        \begin{equation}
                \mathcal{L}(\vec{k} \mid \varepsilon, \vec{\theta}) \, \rightarrow \, \mathcal{L}(\vec{k} \mid \varepsilon, \vec{\theta}) \cdot \prod_\nu \exp \left[-\frac{\left(\theta_\nu-\Phi_\nu^{\mathrm{SSM}}\right)^2}{2\left(\delta_{\Phi_\nu^{\mathrm{SSM}}}\right)^2}\right],
        \end{equation}
        where $\theta_\nu$ represents the floating value of the neutrino flux $\Phi_\nu$, $\Phi_\nu^{\mathrm{SSM}}$ is the expected flux from the SSM prediction. $\delta_{\Phi_\nu^{\mathrm{SSM}}}$ is its uncertainty that results from theoretical uncertainties of the SSM.

\begin{figure}[!hbt]
%\nolinenumbers
    \centering
    \includegraphics[width=0.9\linewidth]{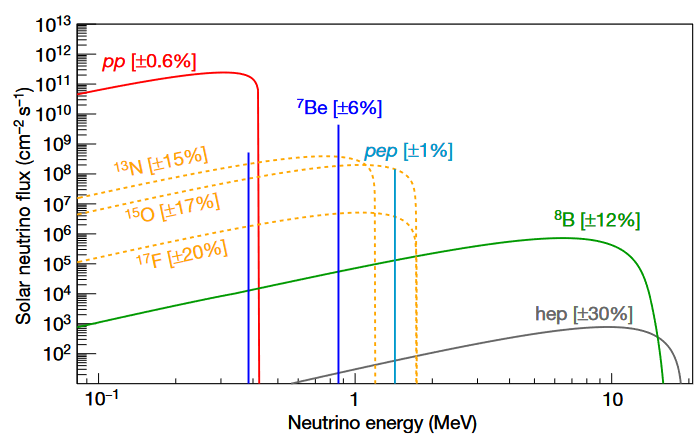}
    \caption{Energy distribution of solar neutrinos produced in the Sun\cite{bib:global}.}
    \label{fig:flux_dist}
%\linenumbers
\end{figure}

\begin{table}[!t]
%\nolinenumbers
    \centering
   \begin{tabular}{ccc}
    \hline\hline Flux, $\Phi_\nu$ & B16(GS98)-HZ & B16(AGSS09met)-LZ \\
    \hline$p p$ & $5.98(1 \pm 0.006)$ & $6.03(1 \pm 0.005)$ \\
    $p e p$ & $1.44(1 \pm 0.01)$ & $1.46(1 \pm 0.009)$ \\
    ${ }^7 \mathrm{Be}$ & $4.93(1 \pm 0.06)$ & $4.50(1 \pm 0.06)$ \\
    $\mathrm{CNO}$ & $4.88(1 \pm 0.11)$ & $3.51(1 \pm 0.10)$ \\
    \hline\hline
    \end{tabular}
    \caption{The fluxes predicted by HZ and LZ-SSM's\cite{bib:SSM}. Units are: $10^{10}(p p), 10^9\left({ }^7 \mathrm{Be}\right), 10^8\left(\right.$ pep, CNO) $\mathrm{cm}^{-2} \mathrm{~s}^{-1}$.}
    \label{tab:ssm_flux}
%\linenumbers
\end{table}
As Tab. \ref{tab:ssm_flux} shows, SSM gives a detailed prediction for solar neutrino fluxes. The energy spectra of the neutrino fluxes predicted by the SSM are reported in Fig.~\ref{fig:flux_dist}. We can observe that $pp, \text{CNO}, ^8\text{B}, hep$ solar neutrinos have a continuous energy spectrum, while $^7\text{Be}, pep$ solar neutrinos are monochromatic. In particular, the latter plays a fundamental role in NSI analysis because the normalization and shape can be well constrained in the fitting process.

 \subsection{Likelihood construction}\label{likelihood}
        To effectively constrain NSI parameters that take advantage of data from Borexino Phase-II and Phase-III, it is imperative to formulate a robust likelihood function that integrates the effects of NSI. The MC-based PDFs of the energy spectra predicted for the Phase-II and Phase-III datasets are produced with \texttt{G4Bx} for each component. The next phase involves including the impact of NSI on these PDFs according to Sec. \ref{sec:teo}. Subsequently, the NSI-induced alterations in the cross section can be convolved with the detector response and resolution (\emph{reweighted}), obtained from the MC simulation, to compute the final PDFs for the spectral analysis.
        The resulting log-likelihood to be maximized reads: 
        \begin{equation}
        \begin{split}
        \ln \mathcal{L}&=\ln \mathcal{L}\left(\varepsilon_{NSI}\right)_{TFC_{sub}}+\ln \mathcal{L}\left(\varepsilon_{NSI}\right)_{TFC_{tag}}\\&+ \ln \mathcal{L}^{SSM}+ \ln \mathcal{L}_{penalty}
        \end{split}
        \end{equation}
        where\\
        \begin{equation}
            \ln \mathcal{L}^{S S M}=\ln \mathcal{L}_{\Phi_{^7{Be}}}^{SSM}+\ln \mathcal{L}_{\Phi_{pep}}^{SSM}+\ln \mathcal{L}_{\mathbf{\Phi}_{CNO}}^{SSM}+\ln \mathcal{L}_{\Phi_{pp}}^{SSM}\, .
        \end{equation} 

Summarizing, as for the previous Borexino analyses, the log-likelihood is built in a way of fitting simultaneously the three-fold coincidence (TFC) tagged and subtracted spectra, while constraining contaminants and solar fluxes as described above. In particular, the term $\mathcal{L}_{\Phi}^{SSM}$ denotes the Gaussian-like likelihood functions derived from the SSM \cite{bib:SSM} penalty. The logarithmic penalty term, $\ln \mathcal{L}_{penalty}$, instead represents the cumulative sum of independent penalties applied to the backgrounds, as detailed in Sec. \ref{penalty}. Table \ref{tab:ssm_flux} presents the predicted fluxes along with their corresponding uncertainties for each solar neutrino component incorporated into our likelihood function. In particular, the flux of $^8\rm{B}$ solar neutrinos is fixed to the SSM prediction, as the Borexino detector lacks sufficient sensitivity to constrain this parameter independently. The comprehensive formulation of the likelihood function has previously been discussed in \cite{bib:global} and \cite{bib:cno}. The cost function to be minimized in our fitting procedure is defined as $-2\ln\mathcal{L}_{total}$, where $\ln\mathcal{L}_{total}$ is the sum of all the terms of the aforementioned likelihood function.  
        
\section{Results} \label{sec:res}

    In order to perform NSI analysis, we preliminarily reproduce the published spectrum fitting results based on Phase-II data. The predicted spectrum is calculated by determining each component's contribution and then summing them up together. Here, the contribution of each component is calculated as follows:
    \begin{equation}
        R \times {\rm exposure} \times \text{PDF}
    \end{equation}
    where $R$ is the rate of each component, which is calculated with solar neutrino flux in the analysis program; $exposure$ is calculated by the effective target mass times the effective data-taking time; $\text{PDF}$ is the probability density distributions along \texttt{nhits} for the corresponding component, which is simulated by \texttt{G4Bx} and \emph{reweighted}. Then we construct the likelihood function using the predicted spectrum and the Phase-II data spectrum, the detailed constructing process has been mentioned in Sec. \ref{likelihood}.

    The fitting strategy of Phase-III spectrum is similar to Phase-II analysis procedure. The differences mainly exist on the penalty and the fitting range, which have been discussed in Sec.~\ref{likelihood}. The reproduced Phase-III spectrum fitting result has been shown in Fig. \ref{fig:P3spectrum}.
    
   The fitting program used demonstrated exceptional performance in independently replicating the Phase-II and Phase-III spectrum analyzes. In order to facilitate concurrent NSI analysis of the Phase-II and Phase-III data, we initially deactivated the NSI and amalgamated the likelihood functions to assess the program's performance in simultaneously fitting both phases of data. The parameters common to the cost functions of Phase II and Phase III are the neutrino fluxes and NSI parameters, as they are assumed to remain constant over time. Upon combining the data spectra from Phase-II and Phase-III, the error of neutrino fluxes decreased, while the mean values for all the fitting parameters remained almost unchanged.
    \begin{figure}[htbp]
    %\nolinenumbers
        \centering
        \includegraphics[width=0.9\columnwidth]{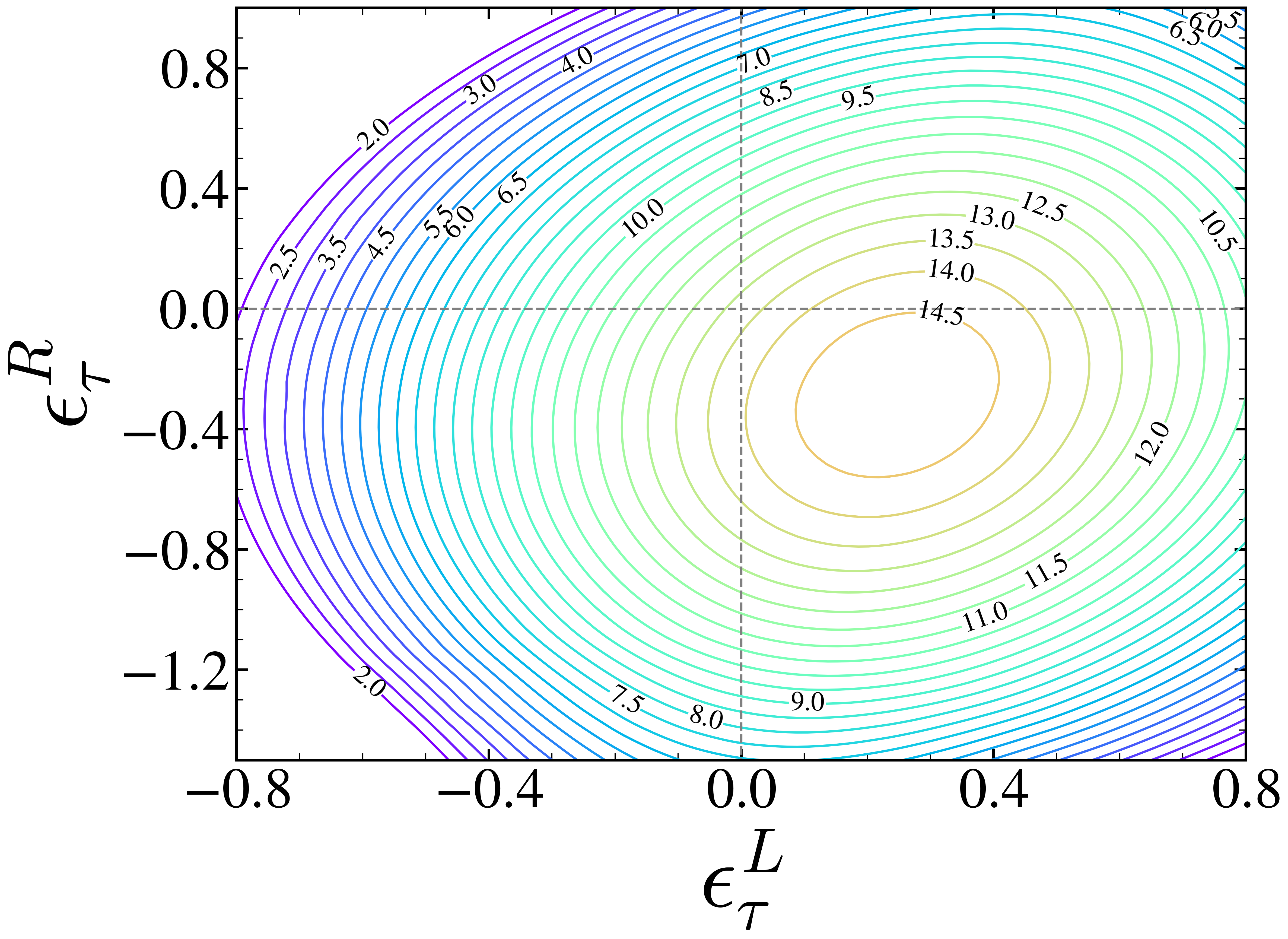}
        \caption{Contour for the rate of $^{210}\rm{Bi}$ on $\varepsilon_{\tau}^{L/R}$ plane with $^{210}\rm{Bi}$ penalty applied. A high $^{210}\rm{Bi}$ rate is preferred at certain region and results in a rejected area of 90\%CL allowed regions.}
        \label{fig:phase3_bi210}
    %\linenumbers
    \end{figure}
   
    \subsection{$\varepsilon^{L/R}$ parameterization}
    \label{LRparameterization}
\begin{figure*}
    %\nolinenumbers
        \centering
        \begin{subfigure}{0.35\linewidth}
        \centering
            \includegraphics[width=\columnwidth]{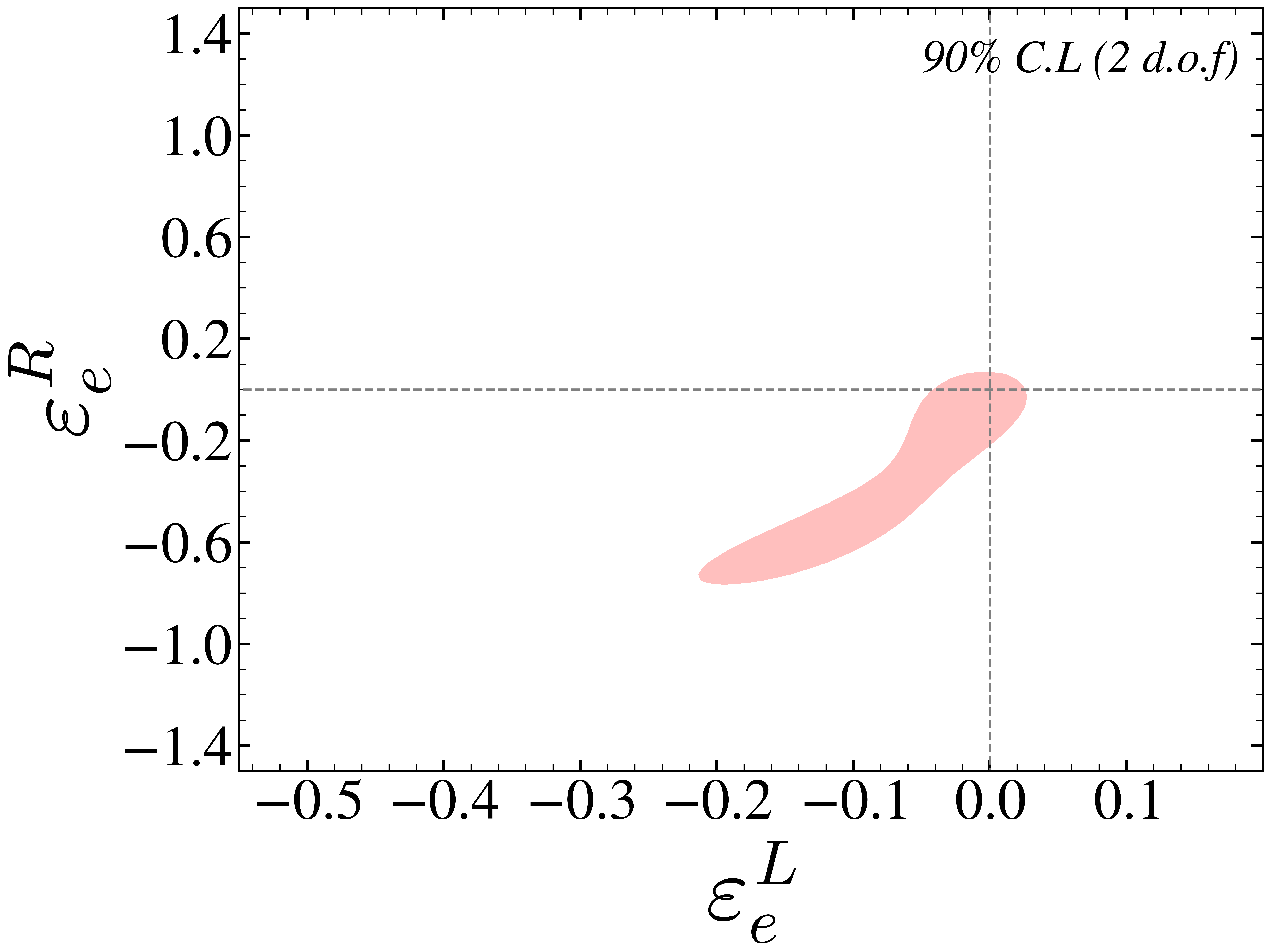}
        \end{subfigure}
        \begin{subfigure}{0.35\linewidth}
            \centering
            \includegraphics[width=\columnwidth]{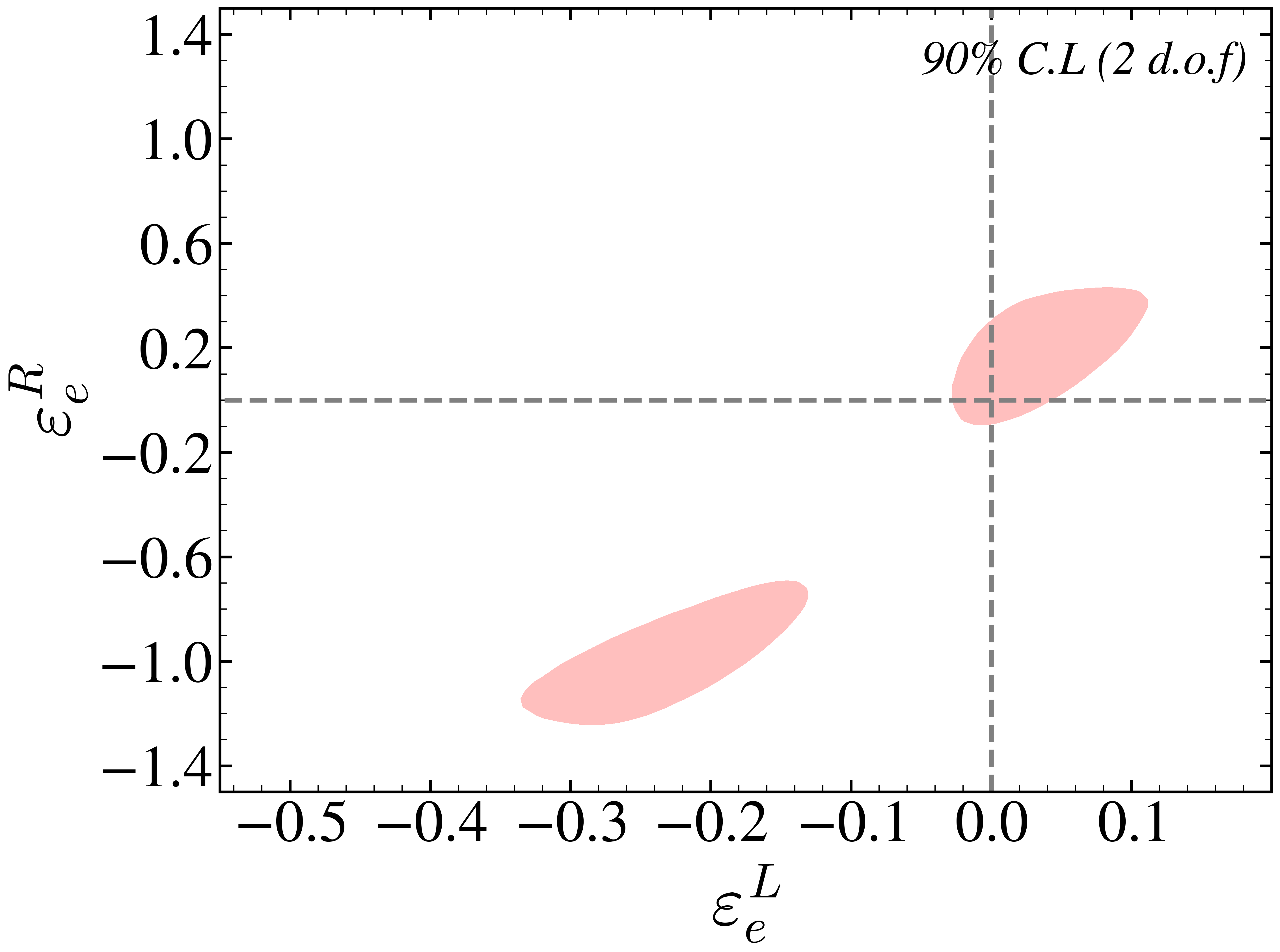}
        \end{subfigure}

        \begin{subfigure}{0.35\linewidth}
        \centering
            \includegraphics[width=\columnwidth]{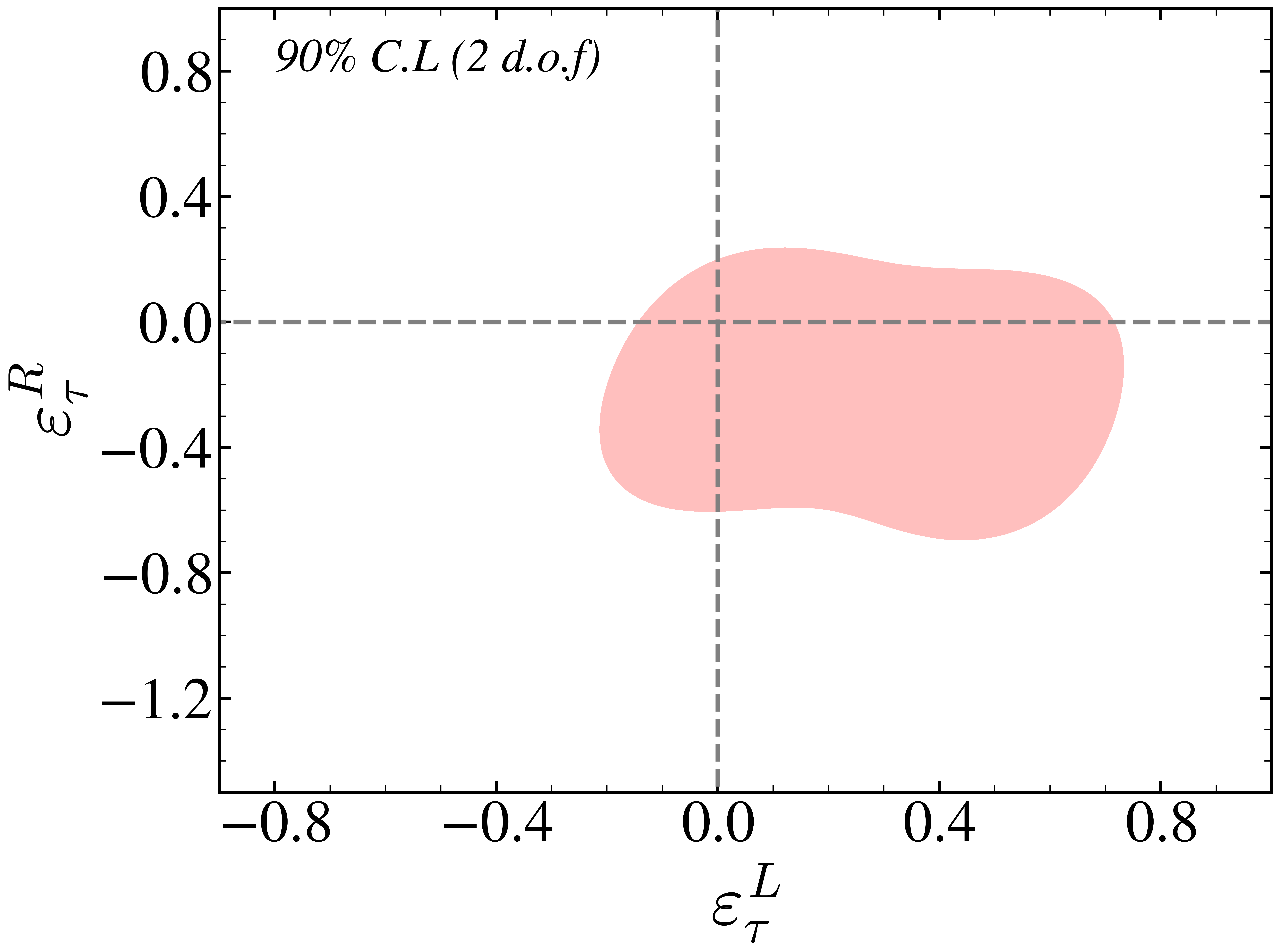}
        \end{subfigure}
        \begin{subfigure}{0.35\linewidth}
            \centering
            \includegraphics[width=\columnwidth]{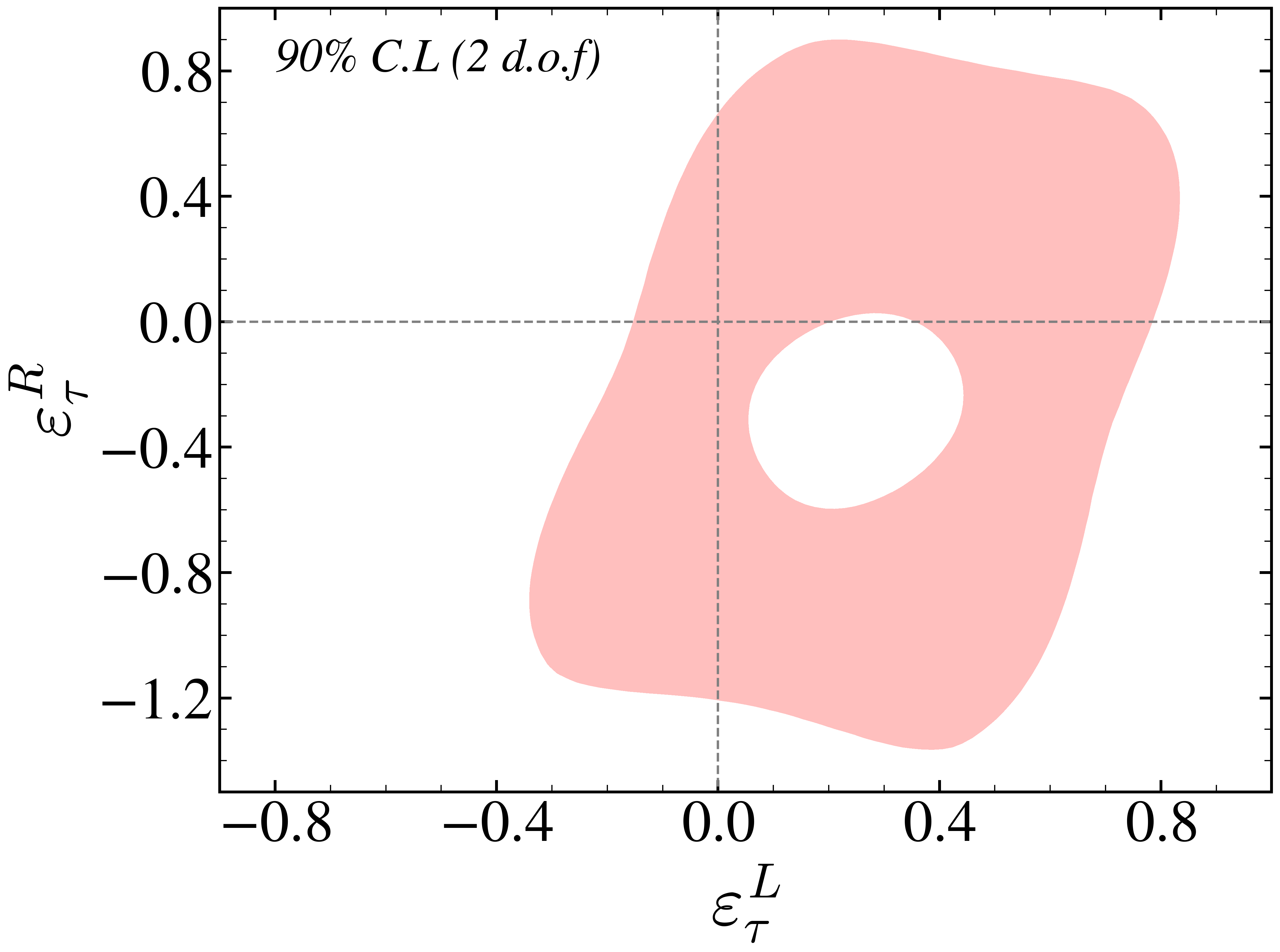}
        \end{subfigure}
        \caption{90\%(CL) allowed regions for NSI parameters in $\varepsilon_{ee}^{L/R}$ and $\varepsilon_{\tau\tau}^{L/R}$ plane. Other NSI parameters are fixed to zero.}
        \label{fig:2Dseperate}
    %\linenumbers
    \end{figure*}

    Fig.~\ref{fig:2Dseperate} illustrates the two-dimensional constraints at a 90\%~CL for the NSI diagonal parameters $\varepsilon_{ee/\tau\tau}^{L/R}$($\varepsilon_{\mu\mu}^{L/R}$ are aleady fixed to 0 from~\cite{bib:charm}) based on the Borexino Phase-II (left) and Phase-III (right) spectral data, all the penalties are applied as illustrated in Sec.~\ref{penalty}. While focusing on NSI diagonal terms, we simplify $\varepsilon_{\alpha\alpha}^{L/R}~(\alpha=e,~\mu,~\tau)$ to $\varepsilon_{\alpha}^{L/R}(\alpha=e,~\mu,~\tau)$ as~\cite{bib:bxnsi} for better comparison. The plots on the left display the reproduced correlated constraints on $\varepsilon_{e}^{L/R}$ and $\varepsilon_{\tau}^{L/R}$ using Phase-II data, while the plots on the right depict the allowed regions for these NSI coefficients based on Phase-III data. Notably, two local minima are observed for $\varepsilon_{e}^{L/R}$, as the combinations of $\varepsilon_{e}^{L/R}$ at these minima exert identical effects on the neutrino-electron elastic scattering cross sections. Phase-III data provide enhanced sensitivity, effectively distinguishing between these two local minima. For the $\varepsilon_{\tau}^{L/R}$ contour using Phase-III data, a forbidden region is enclosed by the 90\%~CL allowed regions due to the penalty applied to $^{210}\rm{Bi}$ in the Phase-III data analysis \cite{bib:cno}. Typically, modifications to the cross section from $\varepsilon_{\tau}^{L/R}$ are counterbalanced by other components during fitting. However, the $^{210}\rm{Bi}$ penalty rejects this compensation, resulting in an increased cost function minimum while the fitting converged, which ultimately manifests itself as a forbidden area within the 90\%CL allowed regions. 

\begin{figure}[htbp]
    %\nolinenumbers
        \centering
        \includegraphics[width=0.9\columnwidth]{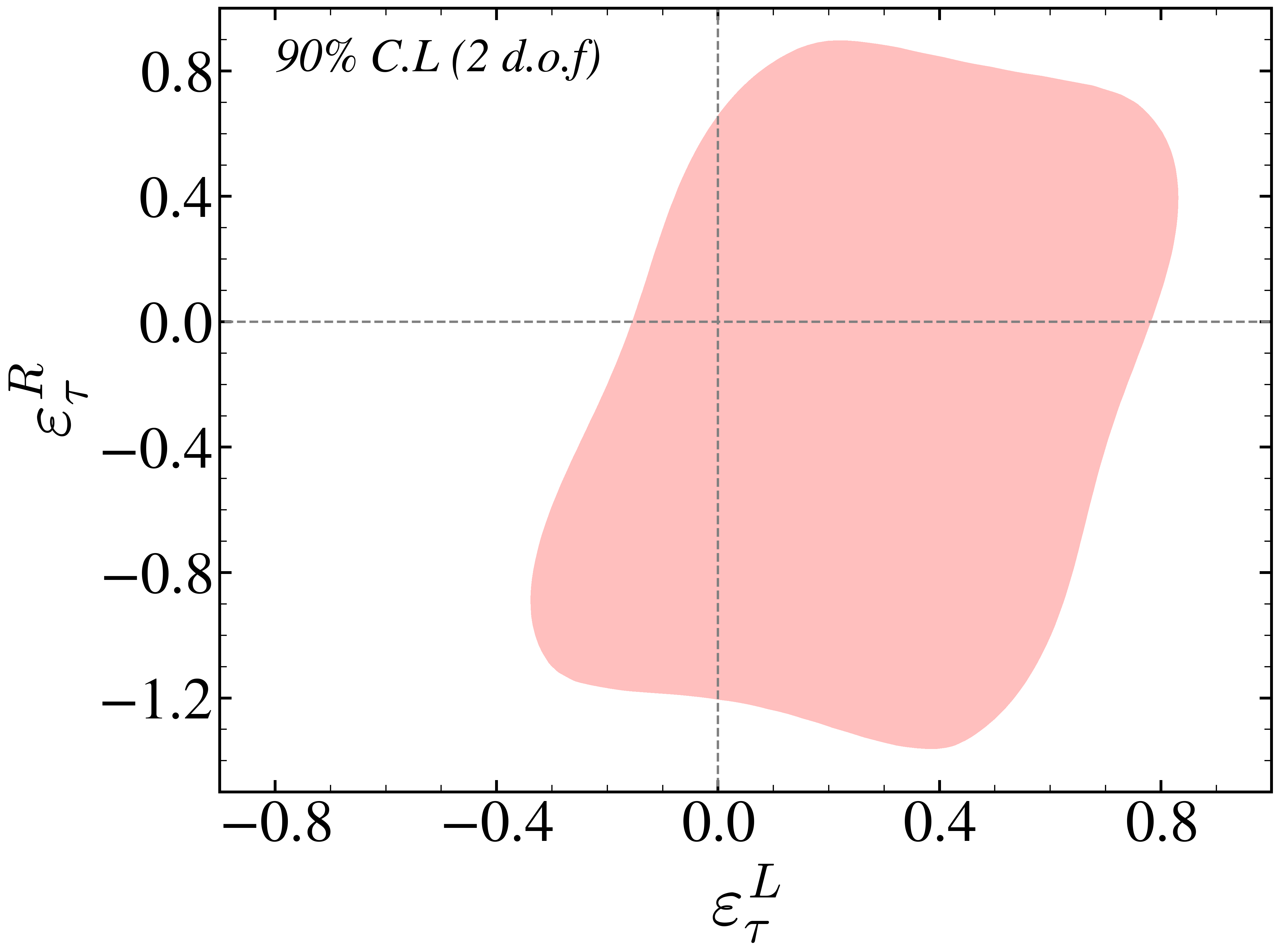}
        \caption{90\% C.L level allowed regions for $\varepsilon_{\tau\tau}^{L/R}$ based on Phase-III data only after removing $^{210}\rm{Bi}$ penalty. The forbidden region has disappeared.}
        \label{fig:phase3_tauLRnobi}
    %\linenumbers
    \end{figure}

Fig.~\ref{fig:phase3_bi210} shows the rate of the $^{210}\rm{Bi}$ contour on the $\varepsilon_{\tau}^{L/R}$ plane. We can observe a preferred high rate of $^{210}\rm{Bi}$ exactly among the forbidden region. After canceling the $^{210}\rm{Bi}$ penalty, the forbidden region in $\varepsilon_{\tau}^{L/R}$ will disappear as Fig. \ref{fig:phase3_tauLRnobi} shows.

\begin{figure*}[htbp]
    %\nolinenumbers
        \centering
        \begin{subfigure}{0.38\linewidth}
        \centering
        \includegraphics[width=\textwidth]{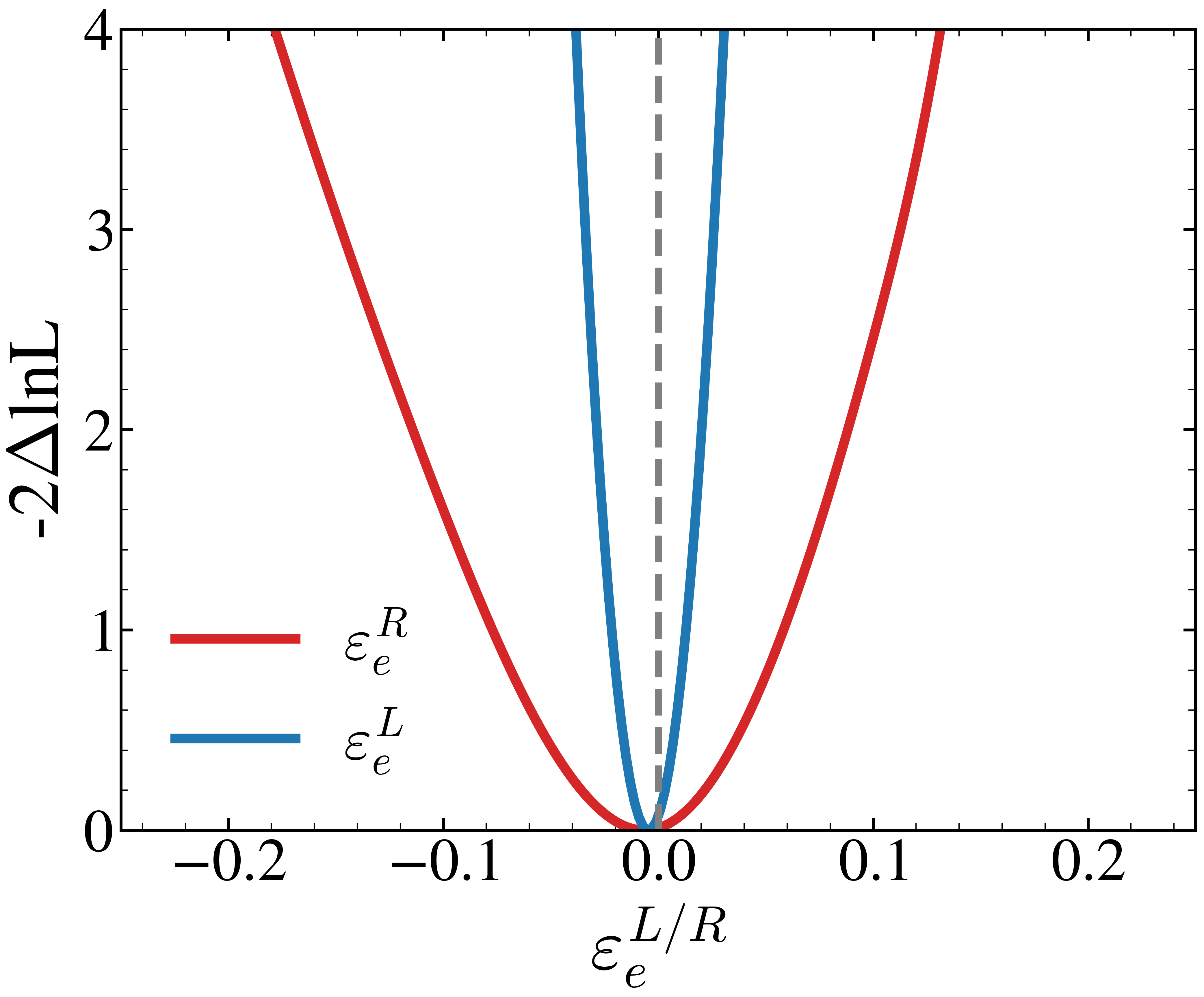}
        \end{subfigure}
        \begin{subfigure}{0.38\linewidth}
        \centering
        \includegraphics[width=\textwidth]{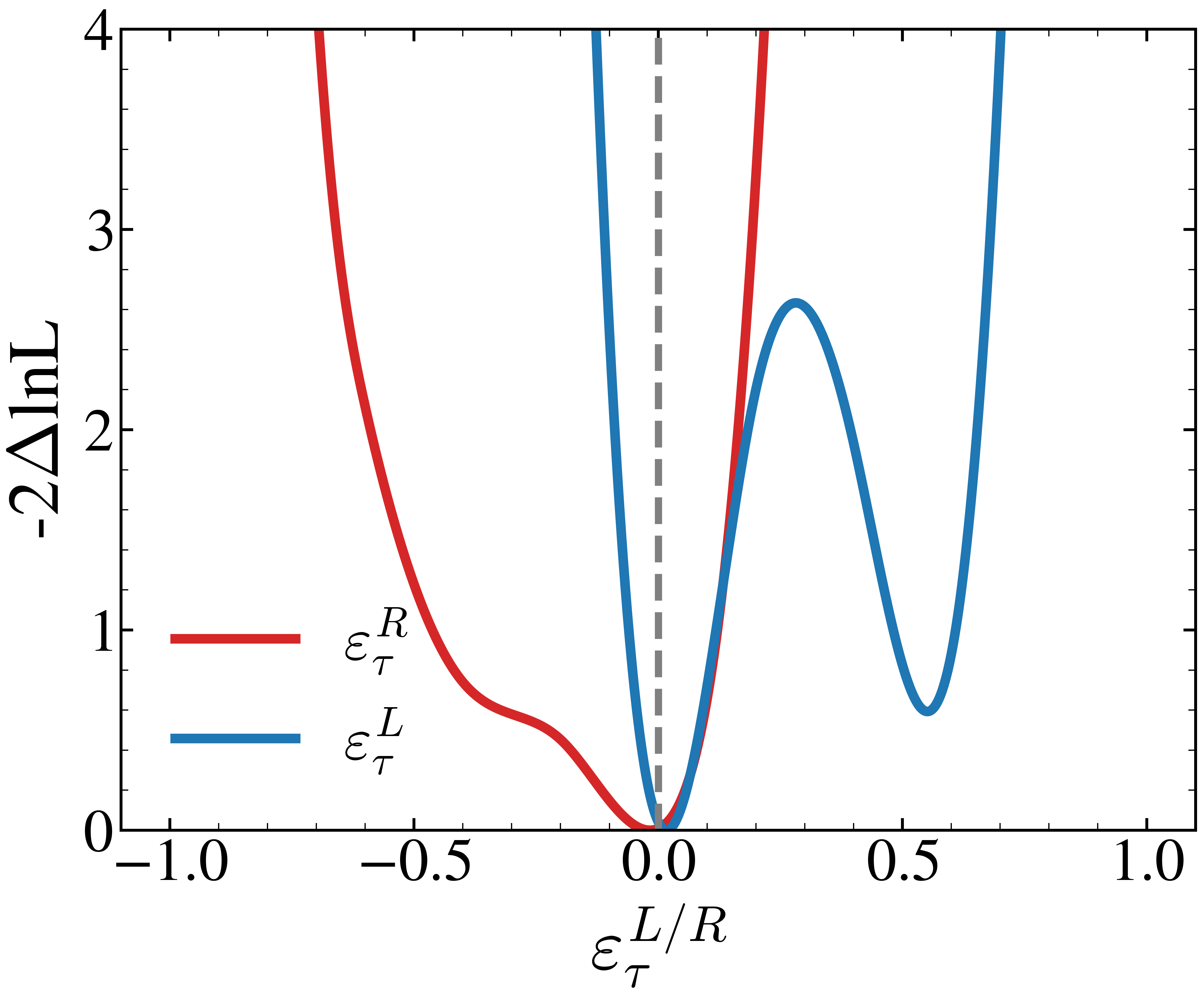}
        \end{subfigure}
        \caption{Left panel shows the log-likelihood profiles for the NSI parameter $\varepsilon_{e}^{L}$(blue line) and $\varepsilon_{e}^{R}$(red line). Right panel shows the same for $\varepsilon_{\tau}^{L}$(blue line) and $\varepsilon_{\tau}^{R}$(red line). Only one NSI parameter were considered at-a-time, the remaining NSI parameters were fixed to zero during profiling.}
        \label{fig:1DprofilingP23}
    %\linenumbers
    \end{figure*}

    Following the same analysis strategy, we can combine Phase-II and Phase-III data by adding up the corresponding likelihood functions together. Fig.~\ref{fig:1DprofilingP23} shows the profiles for the NSI parameter $\varepsilon_{e/\tau}^{L/R}$. After combining, the constrain was improved compared with Phase-II or Phase-III data only.

    \begin{figure*}[htbp]
    %\nolinenumbers
        \centering
        \begin{subfigure}{0.45\linewidth}
        \centering
        \includegraphics[width=\columnwidth]{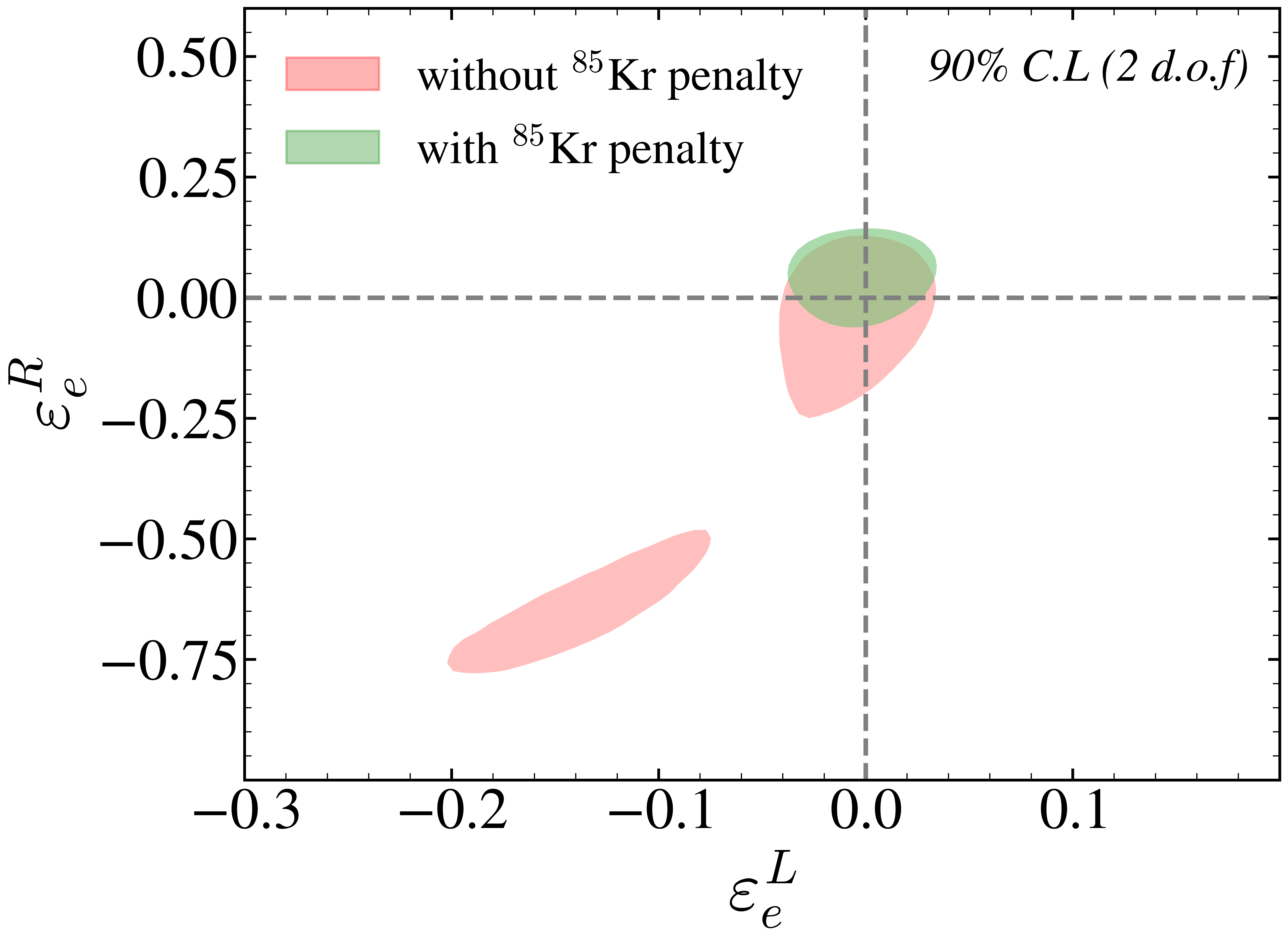}
        \end{subfigure}
        \begin{subfigure}{0.45\linewidth}
        \centering
        \includegraphics[width=\columnwidth]{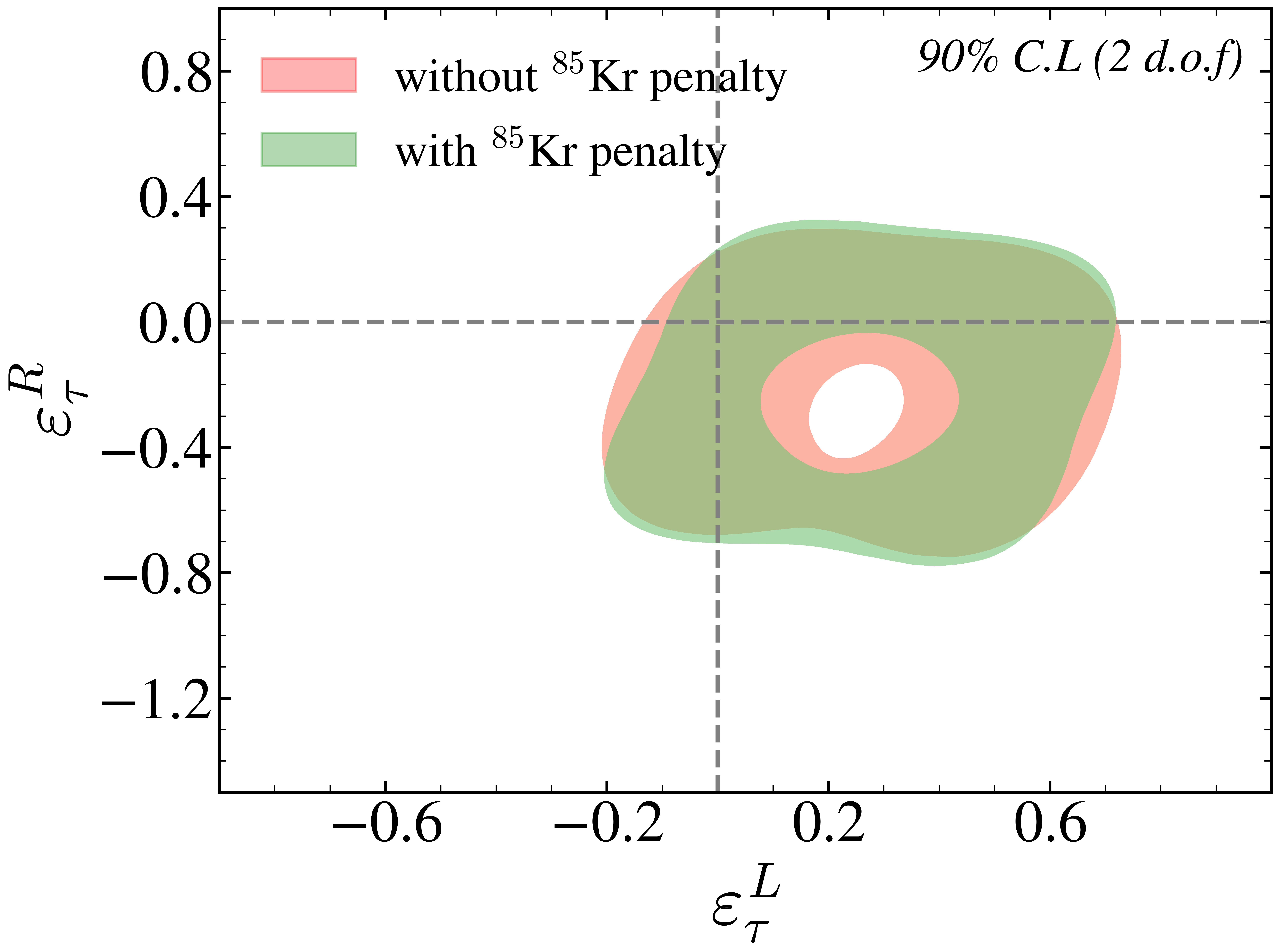}
        \end{subfigure}
        \caption{90\% CL region for two-dimensional contour of $\varepsilon_{e}^{L/R}$ and $\varepsilon_{\tau}^{L/R}$, the red filled area shows contour without $^{85}$Kr penalty and the green filled area shows contour with $^{85}$Kr penalty applied.}
        \label{fig:LR_23_Kr}
    %\linenumbers
    \end{figure*}

    \begin{figure*}[htbp]
    %\nolinenumbers
        \centering
        \begin{subfigure}{0.45\linewidth}
        \centering
        \includegraphics[width=\columnwidth]{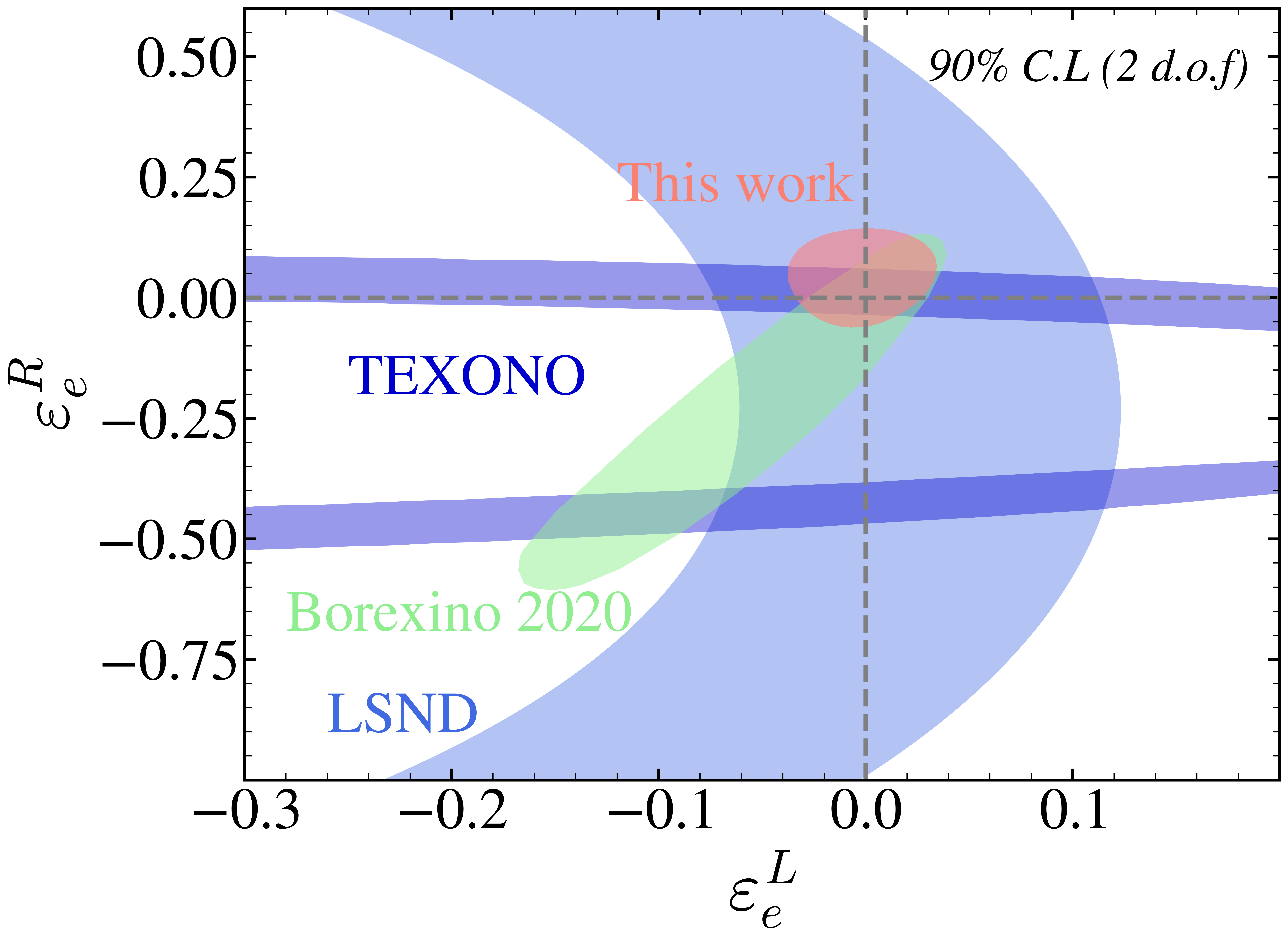}
        \end{subfigure}
        \begin{subfigure}{0.45\linewidth}
        \centering
        \includegraphics[width=\columnwidth]{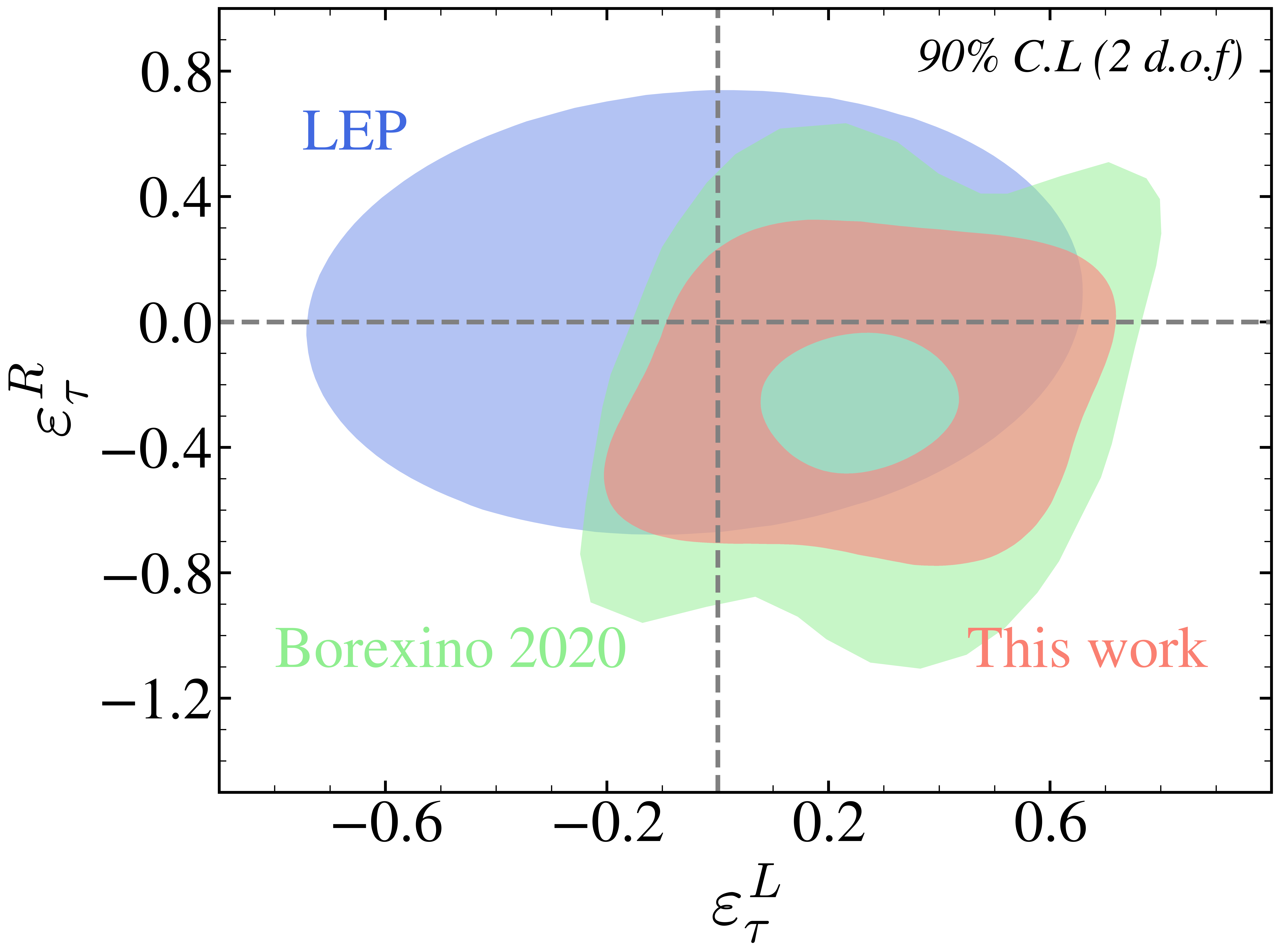}
        \end{subfigure}
        \caption{90\% C.L allowed regions for NSI parameters in $\varepsilon_{e}^{L/R}$(left) and $\varepsilon_{\tau}^{L/R}$(right), other parameters are fixed to zero. Constraints from LSND\cite{bib:LSND} and TEXONO\cite{bib:TEXONO} on $\varepsilon_e^{L/R}$ and Constraints from LEP\cite{bib:LEP} on $\varepsilon_\tau^{L/R}$ are taken for comparison.}
        \label{comparison}
    %\linenumbers
    \end{figure*}
    Fig.~\ref{fig:LR_23_Kr} shows the result for the two-dimensional contour of $\varepsilon_{e}^{L/R}$ and $\varepsilon_{\tau}^{L/R}$ with/without $^{85}$Kr penalty. All the other NSI parameters were fixed to zero for each contour. The red-shadow region shows the 90\% confidence level region for NSI parameters from fitting Borexino Phase-II data only while the green shadow region shows the same result after applying $^{85}\rm{Kr}$ penalty. There is a peaking structure in $\varepsilon_{e}^{L/R}$ 90\% CL region, which means at this peaking region the $\Delta\chi^2$ is larger than 90\% CL region permitted. With $^{85}$Kr penalty applied, the 90\% CL region far from SSM in $\varepsilon_{e}^{L/R}$ contour disappeared. In addition,  the other 90\% CL region are better constrained. For $\varepsilon_{\tau}^{L/R}$ two-dimensional contour, the $^{85}$Kr penalty makes the peak structure 
    slightly larger. In summary, the penalty of $^{85}$Kr helps with constraining NSI parameters using the same fit procedure as used in \cite{bib:bxnsi}.

Following the previously established analysis strategy, the NSI study can be extended to include all diagonal NSI terms. 

\begin{figure*}[!htb]
%\nolinenumbers
        \centering
        \includegraphics[width=0.95\textwidth]{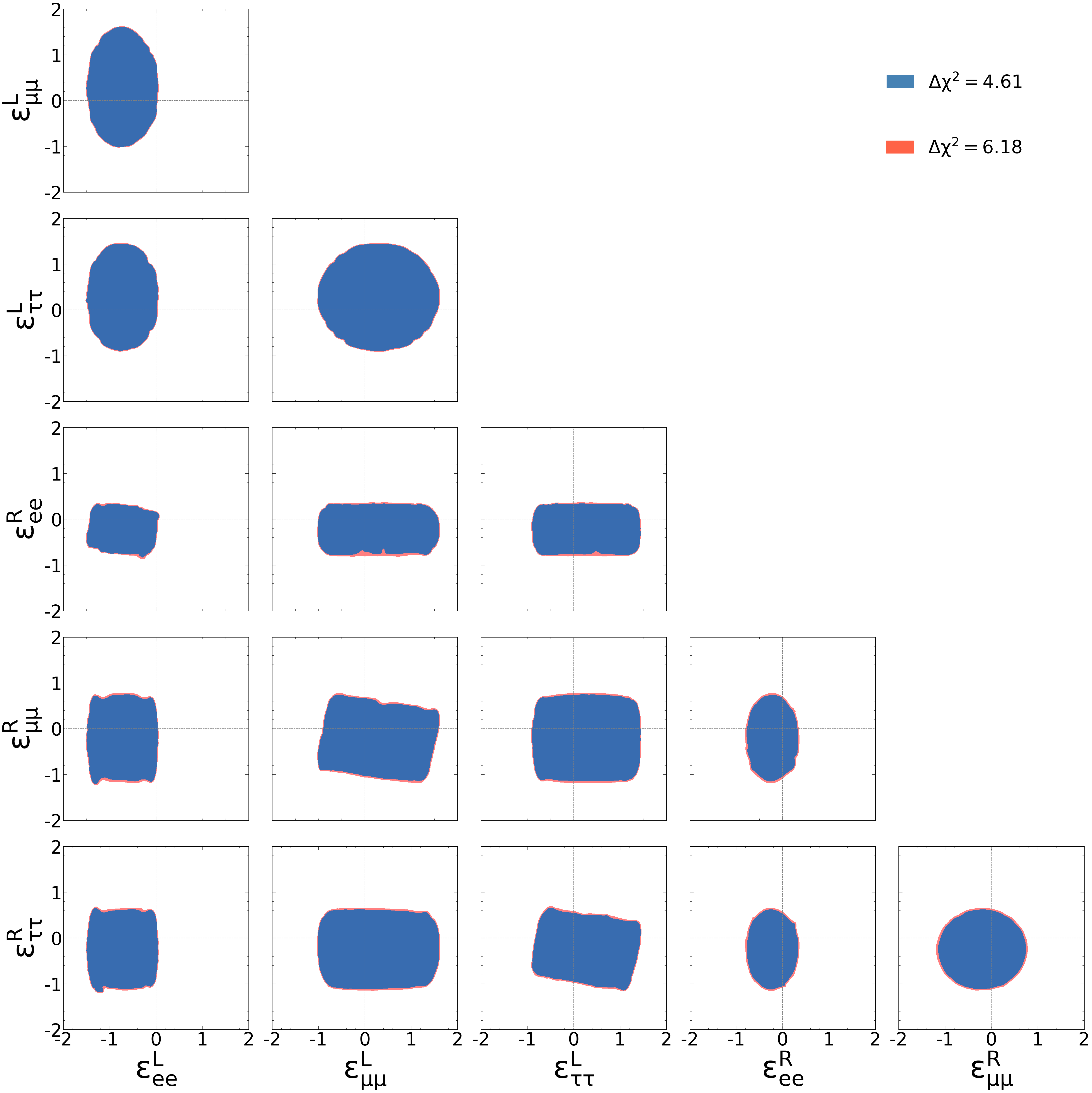}
        \caption{Contours for diagonal NSI parameters $\varepsilon_{\alpha\beta}^{L/R}(\alpha=\beta=e,\mu,\tau)$. Each plot show a two-dimensional allowed confidence regions at 90\% and 95\% (for 2 d.o.f., using two-sided intervals $\Delta\chi^2 = 4.61,\ 6.18$) for different NSI parameter combination. In each panel, the results are obtained after minimizing the cost functions over the other parameters not shown.}
        \label{fig:P23_diagonal}
%\linenumbers
\end{figure*}

Fig.~\ref{fig:P23_diagonal} presents the two-dimensional contours for the diagonal NSI parameters $\varepsilon_{e}^{L/R}, \varepsilon_{\mu}^{L/R}, \varepsilon_{\tau}^{L/R}$. For each contour, all off-diagonal terms are fixed to zero, while the remaining diagonal terms are allowed to float and are marginalized accordingly.

    A hint that the shape of certain contour plots is not an artifact comes from the observation that this (rectangular) deformation appears exclusively in plots comparing left- and right-handed parameters. A similar deformation is also observed when comparing axial and vector couplings with diagonal flavor indices. The origin of this effect can be traced back to the analytical structure of the neutrino–electron scattering cross section. In particular, the cross section contains terms proportional to the product of the left- and right-handed $\varepsilon$ parameters. This can lead to degeneracies in the correlations among the many free parameters, making it challenging to disentangle the individual contributions through marginalization and potentially resulting in distortions of the contour plots.

    Fig.~\ref{comparison} presents the constraints on the parameters $\varepsilon_e^{L/R}$ and $\varepsilon_\tau^{L/R}$ derived from various experiments, including previous results from Borexino \cite{bib:bxnsi} and this work. The LSND experiment exhibits a higher sensitivity to $\varepsilon_e^{L}$, while the TEXONO experiment demonstrates a higher sensitivity to $\varepsilon_e^{R}$. Previous Borexino measurements have already improved the sensitivity to both $\varepsilon_e^{L/R}$, and this work further refines these constraints, resulting in an intersection with the findings of the other three experiments. For $\varepsilon_\tau^{L/R}$, this work also advances beyond the previous Borexino constraints. In particular, it delineates a 90\% confidence level region of rejection, attributed to the $^{210}$Bi penalty effect. This phenomenon has been thoroughly examined in Sec.~\ref{penalty}, providing a comprehensive understanding of its impact on the parameter space. This work strengthens existing constraints and also bridges the gap between different experimental results, offering a more unified and precise understanding of the ranges of allowed parameters for $\varepsilon_e^{L/R}$ and $\varepsilon_\tau^{L/R}$.

\begin{table}[htbp]
%\nolinenumbers
    \centering
    \renewcommand{\arraystretch}{1.5}
    \begin{tabular}{cccc}
    \hline
    \hline
         & 90\% C.L. region &  & 90\% C.L. region\\
        \hline
        $\varepsilon_{e}^{L}$       &    [-0.035, 0.032]~\cite{bib:bxnsi}  & $\varepsilon_{e}^{R}$ &      [0.004, 0.151]~\cite{bib:bxnsi}  \\

        $\varepsilon_{e\mu}^{L}$     &    [-0.17, 0.29]~\cite{Coloma}  & $\varepsilon_{e\mu}^{R}$ &    [-0.21, 0.41]~\cite{Coloma}   \\

        $\varepsilon_{e\tau}^{L}$    &   [-0.26, 0.23]~\cite{Coloma}   & $\varepsilon_{e\tau}^{R}$ &   [-0.35, 0.31]~\cite{Coloma}  \\

        $\varepsilon_{\mu}^{L}$   &    [-0.03, 0.03]~\cite{bib:LEP}   & $\varepsilon_{\mu}^{R}$ &  [-0.03, 0.03]~\cite{bib:LEP}  \\

        $\varepsilon_{\mu\tau}^{L}$  &  [-0.62, -0.52]$\oplus$[-0.09, 0.14]~\cite{Coloma}   & $\varepsilon_{\mu\tau}^{R}$ & [-0.26, 0.23]~\cite{Coloma} \\

        $\varepsilon_{\tau}^{L}$ &    [-0.11, 0.67]~\cite{bib:bxnsi}   & $\varepsilon_{\tau}^{R}$ &[-0.3, 0.4]~\cite{bib:LEP} \\
        \hline
    \end{tabular}
    \caption{Current reported constraints on the NSI coefficients from global experiments.}
    \label{tab:currentlimit_LR}
%\linenumbers
\end{table}

\begin{table}[htbp]
%\nolinenumbers
    \centering
    \renewcommand{\arraystretch}{1.5}
    \begin{tabular}{cccc}
    \hline
    \hline
         & 90\% C.L. region &  & 90\% C.L. region\\
        \hline
        $\varepsilon_{e}^{V}$       &    [-0.09, 0.14]   & $\varepsilon_{e}^{A}$ &      [-0.05, 0.10]   \\

        $\varepsilon_{e\mu}^{V}$     &    [-0.34, 0.61]   & $\varepsilon_{e\mu}^{A}$ &    [-0.30, 0.43]   \\

        $\varepsilon_{e\tau}^{V}$    &    [-0.48, 0.47]   & $\varepsilon_{e\tau}^{A}$ &   [-0.40, 0.38]  \\

        $\varepsilon_{\mu}^{V}$   &    [-0.51, 0.35]   & $\varepsilon_{\mu}^{A}$ &  [-0.29, 0.19]$\oplus$[0.68, 1.45]  \\

        $\varepsilon_{\mu\tau}^{V}$  &    [-0.25, 0.36]   & $\varepsilon_{\mu\tau}^{A}$ & [-1.10, -0.75]$\oplus$[-0.13, 0.22] \\

        $\varepsilon_{\tau}^{V}$ &    [-0.66, 0.52]   & $\varepsilon_{\tau}^{A}$ &[-0.40, 0.36]$\oplus$[0.69, 1.44] \\
        \hline
    \end{tabular}
    \caption{Current reported constraints on $\varepsilon^{V/A}$ parameters from~\cite{Coloma}.}
    \label{tab:currentlimit_VA}
%\linenumbers
\end{table}

Tab.~\ref{tab:currentlimit_LR} summarizes the most stringent constraints currently reported on NSI coefficients. These results are derived under the assumption of varying a single parameter at a time, in contrast to our methodology, which allows other NSI coefficients to fluctuate freely during the constraint of a specific parameter. The strongest bounds on $\varepsilon_{e}^{L/R}$ and $\varepsilon_{\tau}^{L}$ are obtained from Borexino Phase-II data~\cite{bib:bxnsi}, while the most stringent limits on $\varepsilon_{\mu}^{L/R}$ and $\varepsilon_{\tau}^{R}$ originate from a combined analysis of reactor, CHARM II, and LEP data~\cite{bib:bxnsi}. For the off-diagonal terms $\varepsilon_{\alpha\beta}^{L/R}(\alpha\neq\beta)$ and all parameters in the $\varepsilon^{V/A}$ scenario (as summarized in Tab.~\ref{tab:currentlimit_VA}), constraints have been reported exclusively in~\cite{Coloma} using Borexino Phase-II data.

\begin{table}[htbp]
%\nolinenumbers
    \centering
    \renewcommand{\arraystretch}{1.5}
    \begin{tabular}{cccc}
    \hline
    \hline
         & 90\% C.L. region &  & 90\% C.L. region\\
        \hline
        $\varepsilon_{e}^{L}$       &    [-1.47, 0.04]   & $\varepsilon_{e}^{R}$ &      [-0.73, 0.33]   \\
        
        $\varepsilon_{e\mu}^{L}$     &    [-0.63, 0.63]   & $\varepsilon_{e\mu}^{R}$ &    [-0.53, 0.53]   \\
        
        $\varepsilon_{e\tau}^{L}$    &    [-0.62, 0.62]   & $\varepsilon_{e\tau}^{R}$ &   [-0.52, 0.52]  \\
    
        $\varepsilon_{\mu}^{L}$   &    [-1.01, 1.61]   & $\varepsilon_{\mu}^{R}$ &  [-1.13, 0.66]  \\
        
        $\varepsilon_{\mu\tau}^{L}$  &    [-0.84, 0.84]   & $\varepsilon_{\mu\tau}^{R}$ & [-0.63, 0.63] \\
        
        $\varepsilon_{\tau}^{L}$ &    [-0.86, 1.45]   & $\varepsilon_{\tau}^{R}$ &[-1.07, 0.57] \\
        \hline
    \end{tabular}
    \caption{Limits on the NSI coefficients $\varepsilon^{L/R}$ obtained from this work.}
    \label{tab:limit_LR}
%\linenumbers
\end{table}

\begin{figure*}[!htb]
    %\nolinenumbers
        \centering
        \includegraphics[width=0.95\textwidth]{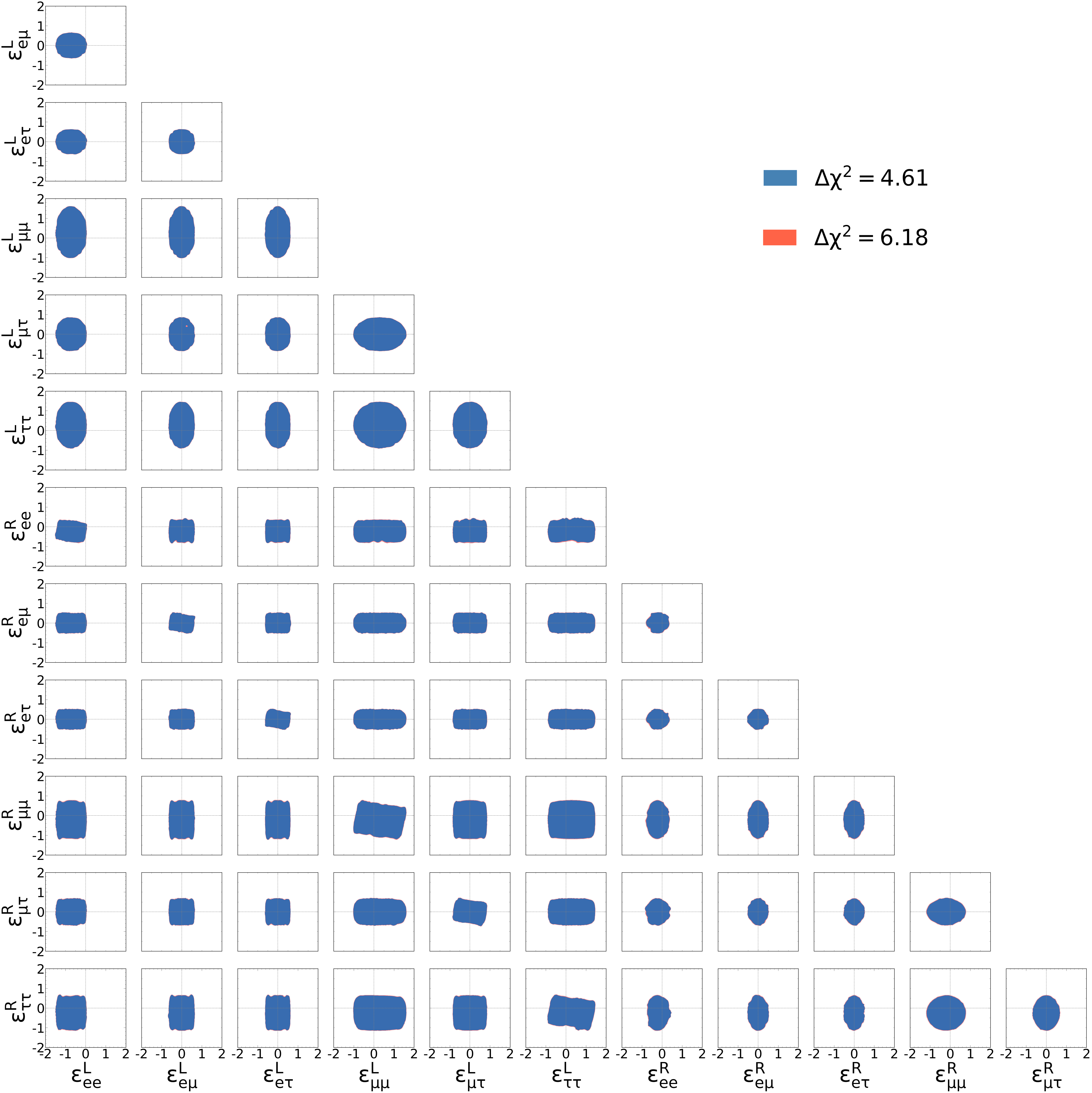}
        \caption{Contours for total 12 NSI parameters $\varepsilon_{\alpha\beta}^{L/R}(\alpha,\beta=e,\mu,\tau)$. The central panels show the two-dimensional allowed confidence regions at 90\% and 95\% (for 2 d.o.f., using two-sided intervals $\Delta\chi^2 = 4.61,\ 6.18$). In each panel, the results are obtained after minimizing the cost functions over the other parameters not shown.}
        \label{allLR_contour}
    %\linenumbers
    \end{figure*}

    Now we can extend our NSI analysis to all 12 NSI parameters, including all the diagonal and off-diagonal terms. Fig.~\ref{allLR_contour} presents the constraints for total 12 NSI coefficients with $\varepsilon^{L/R}$ parameterization.

\subsection{$\varepsilon^{V/A}$ parameterization}
\label{VAparameterization}
For the \(\varepsilon^{V/A}\) parameterization, axial NSI \(\nu\)-e operators are characterized by the property:
$$
\varepsilon_{\alpha \beta}^L = -\varepsilon_{\alpha \beta}^R.
$$

In this scenario $\varepsilon_{\alpha\beta}^{V}=\frac{\varepsilon_{\alpha\beta}^{L}+\varepsilon_{\alpha\beta}^{R}}{2},~\varepsilon_{\alpha\beta}^{A}=\frac{\varepsilon_{\alpha\beta}^{L}-\varepsilon_{\alpha\beta}^{R}}{2}$. The oscillation probability remains unaffected by the introduction of NSI perturbations. This is because the NSI-matter Hamiltonian matrix, which influences the oscillations, is identically zero. This matrix, as described in Eq.~\ref{1} and Eq.~\ref{1a-bis}, is constructed by summing the left and right contributions. Consequently, neutrino oscillations are insensitive to the introduction of axial NSI perturbations. 

The results of the analysis in this context can be represented by comparing different NSI parameters in pairs, after marginalizing the probability over the other parameters not shown. The marginalization process involves integrating the probability over the parameters that are not considered, allowing them to vary freely within the parameter space. The number of free parameters (different neutrino flavor combinations for left and right parameters) can be easily computed due to the symmetry relations, reducing the total to six.
\begin{table}[htbp]
%\nolinenumbers
    \centering
    \renewcommand{\arraystretch}{1.5}
    \begin{tabular}{cccc}
    \hline
    \hline
         & 90\% C.L. region &  & 90\% C.L. region\\
        \hline
        $\varepsilon_{e}^{V}$       &    [-1.97, 0.19]   & $\varepsilon_{e}^{A}$ &      [-1.47, 0.48]   \\

        $\varepsilon_{e\mu}^{V}$     &    [-0.97, 0.97]   & $\varepsilon_{e\mu}^{A}$ &    [-0.94, 0.85]   \\

        $\varepsilon_{e\tau}^{V}$    &    [-0.96, 0.96]   & $\varepsilon_{e\tau}^{A}$ &   [-0.95, 0.83]  \\

        $\varepsilon_{\mu}^{V}$   &    [-1.7, 1.96]   & $\varepsilon_{\mu}^{A}$ &  [-1.96, 1.96]  \\

        $\varepsilon_{\mu\tau}^{V}$  &    [-1.23, 1.23]   & $\varepsilon_{\mu\tau}^{A}$ & [-1.18, 1.18] \\

        $\varepsilon_{\tau}^{V}$ &    [-1.65, 1.67]   & $\varepsilon_{\tau}^{A}$ &[-1.78, 1.78] \\
        \hline
    \end{tabular}
    \caption{Limits on the NSI coefficients $\varepsilon^{V/A}$ obtained from this work.}
    \label{tab:limit_VA}
%\linenumbers
\end{table}

The vectorial NSI operators can be expressed as:
$$
\varepsilon_{\alpha \beta}^L = \varepsilon_{\alpha \beta}^R.
$$
In this condition, NSI can affect both the matter potential felt by neutrinos in propagation and the elastic scattering process in detection. Also in this case, due to symmetry considerations, the number of free NSI parameters is reduced to six.

\begin{figure*}[htbp]
%\nolinenumbers
        \centering
        \begin{subfigure}{0.4\linewidth}
        \centering
        \includegraphics[width=\columnwidth]{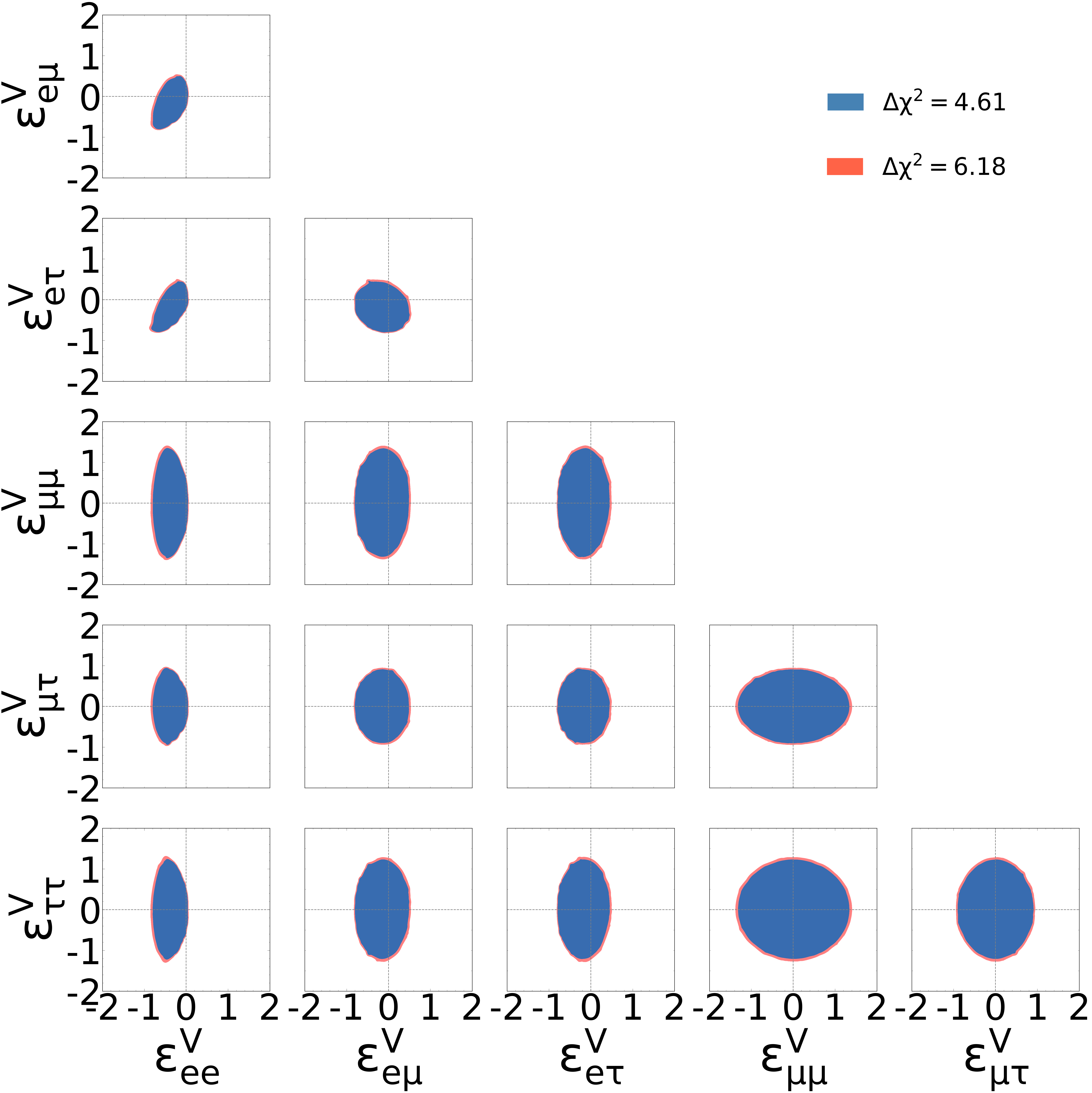}
        \end{subfigure}
        \begin{subfigure}{0.4\linewidth}
        \centering
        \includegraphics[width=\columnwidth]{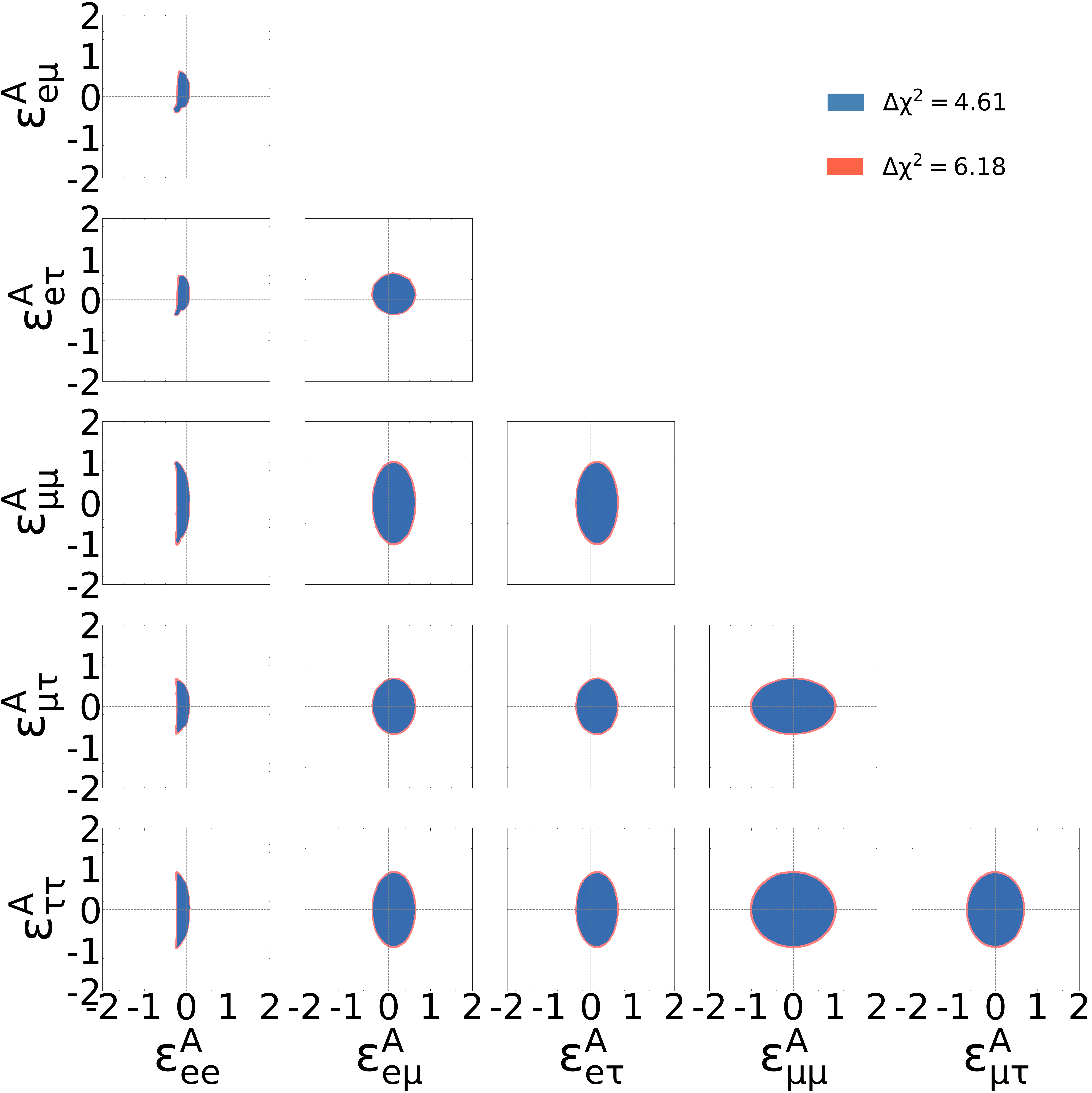}
        \end{subfigure}
        \caption{90\% CL region for two-dimensional contour of $\varepsilon^{A}$(left) and $\varepsilon_{}^{V}$(right) scenario. For each contour, other NSI parameters are freely fluctuating while marginalizing.}
        \label{fig:p23onlyva}
%\linenumbers
\end{figure*}

\begin{figure*}[htbp]
%\nolinenumbers
        \centering
        \includegraphics[width=0.95\textwidth]{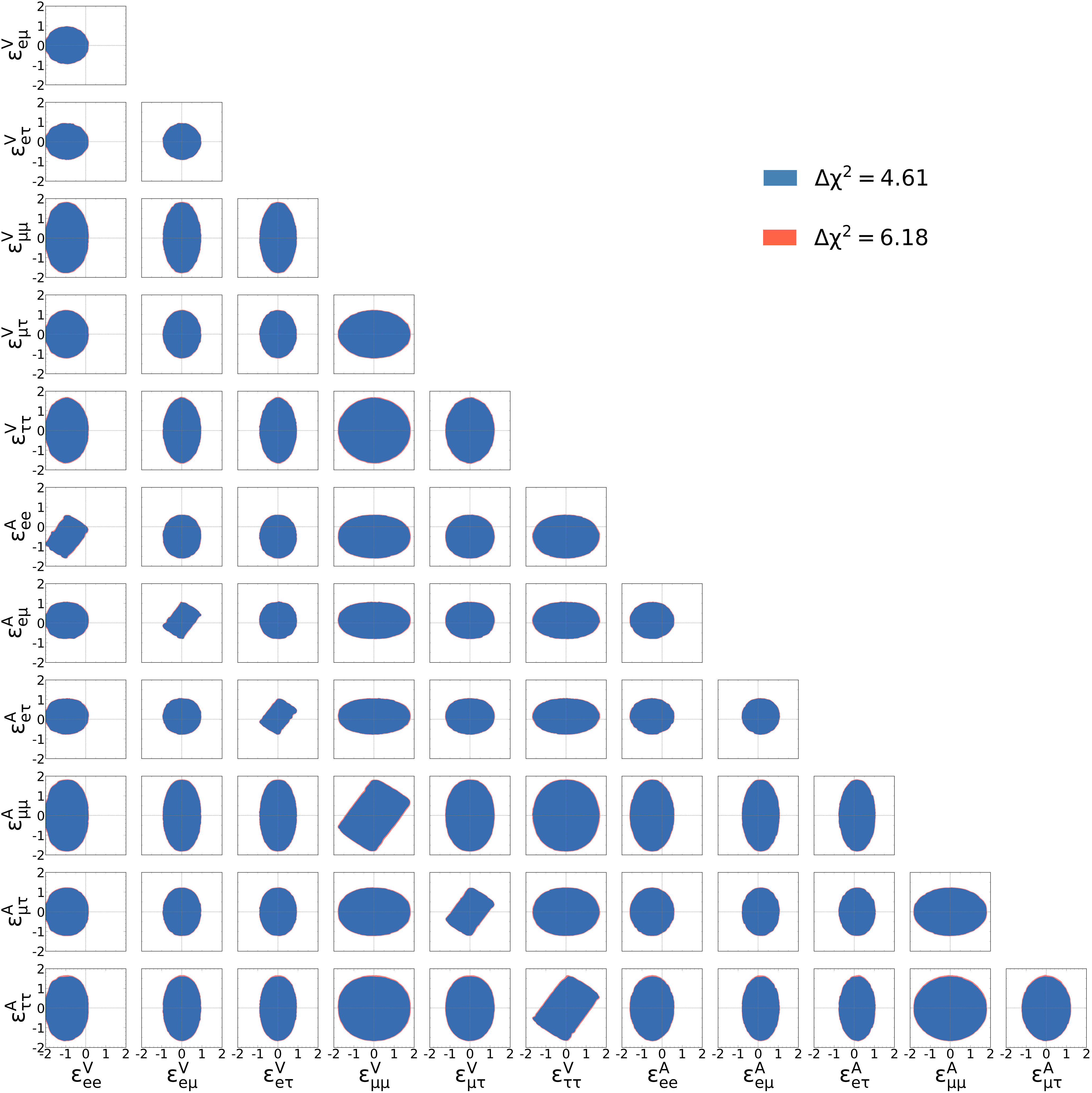}
        \caption{Results for vectorial NSI($\varepsilon_{\alpha\beta}^{V}=\varepsilon_{\alpha\beta}^{L}+\varepsilon_{\alpha\beta}^{R}$), the central panels show the two-dimensional allowed confidence regions at 90\% and 95\% (for 2 d.o.f., using two-sided intervals $\Delta\chi^2 = 4.61,\ 6.18$). In each panel, the results are obtaioned after minimizing the cost functions over the other parameters not shown.}
        \label{vectorial}
%\linenumbers
\end{figure*}

\section*{Conclusions}

In the present work, we search for deviations induced by NSI from the standard neutrino interaction paradigm, generalizing the previous analysis of the Borexino Collaboration~\cite{bib:bxnsi}. Specifically, we investigate non-standard neutrino-electron interactions in the most general scenario, including off-diagonal NSI parameters. Borexino is sensitive to a total of 12 NSI parameters when considering NSI with electrons: six parameters involving left-handed neutrinos interacting with electrons (with coefficients $\varepsilon^{L}_{\alpha\beta}$ ) and six parameters involving right-handed neutrinos (with coefficients $\varepsilon^{R}_{\alpha\beta}$).

NSI are widely explored as a possible portal to new physics beyond the SM, given the peculiar role of neutrinos themselves in motivating the need for more fundamental explanations of their nature and of the logic by which they are classified within the standard paradigm.

Borexino has previously explored NSI couplings in a publication that included the Phase-II dataset~\cite{bib:bxnsi}. These results are now cross-checked using a new  MC-based approach that relies on probability density function \emph{reweighting}. This new method provides greater flexibility and overcomes the challenges of the old analytical approach in Phase-III, where the significant loss of photomultiplier tubes (PMTs) and the resulting decline in detector performance make the analysis more complex. After validating the new analysis tool, we extend the study to the new dataset with increased statistics.

Based on Borexino Phase-II data, the work in ~\cite{bib:bxnsi} has already presented the two-dimensional NSI allowed region for representative cases. In this analysis, the sub-leading effects from the off-diagonal NSI parameters in Eq.~(\ref{cross1}) are ignored and $\varepsilon^{X}_{\mu\mu} (X=L/R)$ is set to zero, following constraints from the $\nu_{e\mu}$  scattering experiment CHARM II~\cite{bib:charm}. Thus, attention is restricted to the remaining four parameters: $\varepsilon^{L/R}_{ee}$ and $\varepsilon^{L/R}_{\tau\tau}$. Furthermore, when plotting the two-dimensional allowed region for $\varepsilon^{R/L}_{ee}$, the parameters $\varepsilon^{L/R}_{\tau\tau}$ are fixed to zero. Fig.~\ref{comparison} presents the allowed region for NSI parameters in the $\varepsilon^{L/R}_{ee}$ plane, following the same analysis strategy as in [\ref{}] after combining Borexino Phase-II and Phase-III data.

As Eq.~(\ref{cross1}) suggests, $\varepsilon^{L}_{\alpha\beta}$ mainly modifies the normalization of neutrino events, while $\varepsilon^{R}_{\alpha\beta}$ mainly influences the shape of the event spectra. Consequently, Borexino spectral data provide information to constrain NSI when fitting with an NSI-included theory. The overall constraints on diagonal terms are further improved compared to the previous analysis, and the strong dependence on independent constraints from $^{85}$Kr and $^{210}$Bi is extensively discussed. In particular, the levels of these contaminants are determined more accurately by incorporating the full dataset, leading to reduced statistical uncertainties.

Finally, for the first time, we present a full analysis that includes all off-diagonal NSI parameters (six left-handed and six right-handed). Moreover, after redefining the NSI parameters, we consider six vector and six axial parameters, obtained as linear combinations of the left- and right-handed ones. The full analysis is conducted by considering pairs of parameters while allowing the others to float in the parameter space.

Despite the larger allowed regions, the Borexino spectral fit is able to accommodate them simultaneously, showing no substantial or significant deviation from their null values.

\section*{Acknowledgments}

The Borexino program is made possible by funding from Istituto Nazionale di Fisica Nucleare
(INFN) (Italy), National Science Foundation (NSF)
(USA), Russian Science Foundation (RSF) (Grant
No. 24-12-00046), Deutsche Forschungsgemeinschaft (DFG), Cluster of Excellence PRISMA+(Project ID 39083149), and recruitment initiative
of Helmholtz-Gemeinschaft (HGF) (Germany), and
Narodowe Centrum Nauki (NCN) (Grant No. UMO2013/10/E/ST2/00180) (Poland). We acknowledge the
generous hospitality and support of the Laboratori
Nazionali del Gran Sasso (Italy).

%\newpage

\appendix
{\footnotesize

\section*{Appendix}
\section{Pontecorvo-Maki-Nacagawa-Sakata (PMNS) matrix}
The PMNS matrix can be expressed in terms of the mixing angles governing the vacuum oscillation pattern:
\begin{widetext}
\begin{equation}
\begin{split}
\label{U}
U&=e^{i\theta_{23} \lambda_{7}} \Delta e^{i\theta_{13} \lambda_{5}} e^{i\theta_{12} \lambda_{2}}=\\
=&\left(
    \begin{array}{ccc}
      1 & 0 & 0 \\
      0 & c_{23} & s_{23} \\
      0 & -s_{23} & c_{23} \\
    \end{array}
  \right)
  \left(
    \begin{array}{ccc}
      1 & 0 & 0 \\
      0 & 1 & 0 \\
      0 & 0 & e^{i\delta} \\
    \end{array}
  \right)
  \left(
    \begin{array}{ccc}
      c_{13} & 0 & s_{13} \\
      0 & 1 & 0 \\
      -s_{13} & 0 & c_{13} \\
    \end{array}
  \right)
  \left(
    \begin{array}{ccc}
      c_{12} & s_{12} & 0 \\
      -s_{12} & c_{12} & 0 \\
      0 & 0 & 1 \\
    \end{array}
  \right)
\end{split}
\end{equation}
\end{widetext}

where $\lambda_{i}$ are the Gell-Mann matrices, $\Delta$ is the CP violation phase matrix, $c_{ij}=\cos\theta_{ij}$ and $s_{ij}=\sin\theta_{ij}$. The three different rotation matrices are usually named $R_{23}=e^{i\theta_{23}\lambda_{7}}$, $R_{13}=e^{i\theta_{13}\lambda_{5}}$ and $R_{12}=e^{i\theta_{12}\lambda_{2}}$.\\

\section{NSI modified solar neutrino oscillation probability}
\label{appendixA}
In the Hamiltonian picture the evolution of a neutrino state is described by the Schr\"oedinger equation
In the case of solar neutrinos, the oscillation probability can be obtained by considering the propagation through solar matter, then through vacuum, and finally through terrestrial matter.
The oscillation probability is obtained as the squared of the transition amplitude:
\begin{equation}
\begin{split}
&P_{\alpha\beta}=|A_{\alpha\beta}|^2=
|\langle\nu_{\beta}|\nu_{\alpha}(t)\rangle|^2=|A_{\alpha\beta}|^2=\\
&=\bigg|\sum_{k}(A^{E}_{\alpha k})^{*}\exp{\left(-i\frac{m_{k}^{2}}{2E_{k}}L\right)}A^{S}_{k\beta}\bigg|^2
\end{split}
\end{equation}
where $A^{E}_{\alpha k}$ is the transition amplitude for a neutrino vacuum mass eigenstate $k$ to a flavor $\alpha$ inside Earth, whereas $A^{S}_{k\beta}$ is the same inside the Sun, that is the transition amplitude for a $\beta$ flavor eigenstate in a $k$ vacuum one. The greek index $\beta$ indicates the sum over the 3 flavor eigenstates and the latin index $k$ stands for the sum over the corresponding mass eigenstates.\\
\begin{equation}
A_{\alpha\beta}=\langle\nu_{\beta}|\nu_{\alpha}(L)\rangle=\sum_{\tau}\sum_{k}U^{*}_{\beta k}e^{-i\frac{m_{k}^{2}}{2E_{k}}L}U_{\alpha k}
\end{equation}

Considering that neutrinos are produced in the electronic flavor inside the Sun, the final form of the electronic neutrino survival probability becomes:
\begin{equation}
\begin{split}
P_{ee}&|A_{ee}|^2=\sum_{k}|A^{E}_{ek}|^2|A^{S}_{ek}|^2\\
&+2\sum_{k>j}A^{E}_{ek}A^{E\dagger}_{ek}A^{S}_{ej}A^{S\dagger}_{ej}\cos\left({\frac{\Delta m_{kj}^{2}}{2E_{\nu}}L+\delta}\right)
\end{split}
\end{equation}
where $\dagger$ is the hermitian adjoint operator and $\delta$ is the usual CP violation phase. The convention that the index $E$ indicates the term relative to the Earth and $S$ indicates the term relative to the Sun is still valid. The last term of the previous equations can be neglected, in fact, the term oscillates rapidly and its contribution is null when averaged over realistic experimental energy bins with finite size. Hence the survival probability becomes:
\begin{equation}
P_{ee}=|A_{ee}|^2=\sum_{k}|A^{E}_{ek}|^2|A^{S}_{ek}|^2=\sum_{k}P^{E}_{ek}P^{S}_{ek}.
\end{equation}
Taking into account the explicit form of the transition amplitudes and neglecting the fast oscillation term for the same motivation introduced before, the survival probability can be written as:
\begin{equation}
\label{survivalprobability}
P_{ee}=\sum_{j}|U^{E}_{ej}|^2|U^{S}_{ej}|^2
\end{equation}
The PMNS matrices used in the previous equality are obtained in the context of the NSI modified MSW effect. The propagation of  neutrinos inside matter is ruled by the Hamiltonian \ref{nsihamiltstandard} where the matter interaction operator is given by:
\begin{equation}
\begin{split}
&H=\;
\frac{1}{2E_{\nu}}\bigg[U\left(
                      \begin{array}{ccc}
                          0 & 0 & 0 \\
                          0 & \Delta m_{21}^2 & 0 \\
                          0 & 0 & \Delta m_{32}^2 \\
                      \end{array}
                   \right)U^{\dagger}\\
&+V_{matt}
\left(
  \begin{array}{ccc}
    1+\varepsilon_{ee} & \varepsilon_{e\mu}e^{i\phi_{e\mu}} & \varepsilon_{e\tau}e^{i\phi_{e\tau}} \\
    \varepsilon_{e\mu}e^{-i\phi_{e \mu}} & \varepsilon_{\mu\mu} & \varepsilon_{\mu\tau}e^{i\phi_{\mu\tau}} \\
    \varepsilon_{e\tau}e^{-i\phi_{e\tau}} & \varepsilon_{\mu\tau}e^{-i\phi_{\mu\tau}} & \varepsilon_{\tau\tau} \\
  \end{array}
\right)\bigg]
\end{split}
\end{equation}
where the matter potential is $V_{matt}=2\sqrt{2}G_{F}N_{e}E_{\nu}$.
The electronic density inside the Sun and the Earth can be approximated as constant, resulting in a value for the matter potential given by:
\begin{equation}
V_{matt}\simeq
\begin{cases}
(4\div8)\times10^{-5}\mathrm{eV}^{2}\left(\mathrm{\frac{E_{\nu}}{1MeV}}\right)\qquad \mathrm{Solar\; matter\; density}\\
(1.5\div2.5)\times10^{-7}\mathrm{eV}^{2}\left(\mathrm{\frac{E_{\nu}}{1MeV}}\right)\;\, \mathrm{Earth\; matter\; density}.
\end{cases}
\end{equation}
To continue the study the change of the neutrino basis using the unitary matrix $O=R_{23}R_{13}$ is useful. This approach allows for the flavor basis reintroduction, resulting in:
\begin{equation}
\widetilde{H}=O^{\dagger}\,H\,O=(R_{23}R_{13})^{\dagger}\,H\,(R_{23}R_{13}).
\end{equation}
The form of the obtained Hamiltonian operator can be simplified considering that the $\theta_{13}$ mixing angle is negligible compared to the other ones. After subtracting a common phase term in the form of the diagonal matrix $\varepsilon_{\tau\tau}V_{matt}\,\mathbb{I}$, taking into account that $\Delta m_{31}^{2}\gg V_{matt}$ for both the solar and terrestrial matter densities and considering $\theta_{13}\simeq0$, the Hamiltonian can be approximated in the form:
\begin{equation}
\widetilde{H}=\left(
                \begin{array}{cc}
                  \widetilde{H}_{2\times2} & 0 \\
                  0 & \frac{1}{2E_{\nu}}\Delta m_{31}^2 + O(\varepsilon\times V_{matt}) \\
                \end{array}
              \right).
\end{equation}
The term $O(\varepsilon\times V_{matt})$ is negligible, as a consequence the third mass eigenstate decouples from the other two and propagate without interacting as in vacuum. Therefore, in the NSI scenario as well, the analysis of the MSW effect reduces to the two flavor eigenstates approximation. As a result, the obtained picture uses the solar neutrino flavor basis, with the third mass eigenstate decoupled. The reduced Hamiltonian can be written in the explicit form:
\begin{equation}
\label{redham1}
\begin{split}
&\widetilde{H}_{2\times2}=\frac{\Delta m_{21}^2}{4E_{\nu}}\left(
\begin{array}{cc}
-\cos{2\theta_{12}} & \sin{2\theta_{12}} \\
\sin{2\theta_{12}} & \cos{2\theta_{12}}
\end{array}
\right)\\
&+\frac{V_{matt}}{2E_{\nu}}\left[
\left(
\begin{array}{cc}
1 & 0\\
0 & 0
\end{array}
\right)+
\left(
\begin{array}{cc}
\widetilde{\varepsilon}_{1} & \widetilde{\varepsilon}_{2}\\
\widetilde{\varepsilon}_{2} & \widetilde{\varepsilon}_{1}
\end{array}
\right)
\right].
\end{split}
\end{equation}
The terms $\widetilde{\varepsilon}$ are the NSI introduced perturbations and are functions of the NSI parameters $\varepsilon_{**}$ and of the mixing angles. After subtracting a common phase term, the $2\times2$ Hamiltonian can be written as:
\begin{equation}
\label{redham2}
\widetilde{H}_{2\times2}\simeq\frac{1}{2E_{\nu}}\widetilde{M}
\end{equation}
with the matrix $M$ defined as:
\begin{widetext}
\begin{equation}
\widetilde{M}=
\left(
\begin{pmatrix}
-\Delta m_{21}^2 \cos{(2\theta_{12})}+2\,V_{matt}\,(1+\varepsilon_{ee}-\varepsilon_{\tau\tau}) & -\Delta m_{21}^2 \sin{(2\theta_{12})}+2\,V_{matt}\cdot\varepsilon_{e\mu}\cdot\cos{\theta_{23}} \\
-\Delta m_{21}^2 \sin{(2\theta_{12})}+2\,V_{matt}\cdot\varepsilon_{e\mu}\cdot\cos{\theta_{23}} & \Delta m_{21}^2 \cos{(2\theta_{12})}+V_{matt}\cdot(\varepsilon_{\mu\mu}-\varepsilon_{\tau\tau})\,\cos^{2}{\theta_{23}}
\end{pmatrix}
\right)
\end{equation}
\end{widetext}
Finally using the MSW standard analysis strategy it is possible to obtain the PMNS matrix that connects the flavor eigenstates with the matter mass ones:
\begin{equation}
\label{matrix}
U_{M}=\left(
        \begin{array}{cc}
          \cos{\theta_{M}} & \sin{\theta_{M}} \\
          -\sin{\theta_{M}} & \cos{\theta_{M}} \\
        \end{array}
      \right)
\end{equation}
with the matter mixing angle given by:
\begin{equation}
\tan{(2\theta_{M})}=\frac{\Delta m_{21}^2\sin{(2\theta_{12})-2\,V_{matt}\cdot\varepsilon_{e\mu}\cdot\cos{\theta_{23}}}}{\Delta m_{21}^2\cos{(2\theta_{12})}-V_{matt}\,(1-\varepsilon')}
\end{equation}
and this result can be written in the final form \ref{thetaM}:
\begin{equation}
\begin{split}
\tan{(2\theta_{M})}=&
\tan{(2\theta_{12})}\bigg[1-\frac{2\,V_{matt}\cdot\varepsilon_{e\mu}\cdot\cos{\theta_{23}}}{\Delta m_{12}^{2}\sin{(2\theta_{12})}}\bigg]\\
&\bigg[1-\frac{V_{matt}(1-\varepsilon')}{\Delta m_{12}^{2}\cos{(2\theta_{12})}}\bigg]^{-1}
\end{split}
\end{equation}
where we have introduced the parameter $\varepsilon'=-\varepsilon_{ee}+\varepsilon_{\mu\mu}\cos^{2}{\theta_{23}}+\varepsilon_{\tau\tau}\sin^{2}{\theta_{23}}$ \ref{epsilon}. The Solar or Earth matter mixing angles with the related PMNS matrix can be obtained choosing the correct matter density for the interaction potential in the previous analysis framework. 

Matter-induced effects can be included considering the matter mixing angle and the related unitary matrix \ref{matrix}. The oscillation probability \ref{survivalprobability} for day neutrinos is obtained neglecting the contribution of Earth matter:
\begin{align}
\label{oscillationprob}
P_{ee}=&\sum_{j}|U^{E}_{ej}|^2|U^{S}_{ej}|^2=\sin^4\theta_{13}+\cos^4\theta_{13}
P_{ee}^{2\nu}(\theta_{M},\,\theta_{12})\notag\\
P_{e\mu}=&\sum_{j}|U^{E}_{\mu j}|^2|U^{S}_{ej}|^2=\cos^2\theta_{13}\sin^2\theta_{23}
\left(1-P_{ee}^{2\nu}(\theta_{M},\,\theta_{12}) \right)\notag\\
P_{e\tau}=&\sum_{j}|U^{E}_{\tau j}|^2|U^{S}_{ej}|^2=\cos^2\theta_{13}\cos^2\theta_{23}
\left(1-P_{ee}^{2\nu}(\theta_{M},\,\theta_{12}) \right)
\end{align}
where $\theta_{M}$ is the effective mixing angle in solar matter, and $P_{ee}^{2\nu}(\theta_{M},\,\theta_{12})$ is the adiabatic two-flavor survival probability for electronic neutrinos. Under the hypothesis of adiabatic MSW effect and in the solar neutrino energy range it can be approximated in the form: 
\begin{equation}
P_{ee}^{2\nu}(\theta_{M},\,\theta_{12})=\frac{1}{2}\left(1+\cos{(2\theta_{M})}\cos{(2\theta_{12})}\right).
\end{equation}
The explicit form of the survival probability, for instance, can be written as:
\begin{equation}
P_{ee}=\sin^4{\theta_{13}}+\frac{1}{2}\cos^4{\theta_{13}}\left(1+\cos{(2\theta_{M-S})}\cos{(2\theta_{12})}\right).
\end{equation}
It is important to underline that the NSI modified survival probability, obtained before, differs from the standard one for perturbative terms:
\begin{equation}
P_{ee}=\sin^4{\theta_{13}}+\frac{1}{2}\cos^4{\theta_{13}}\left(1+\cos^2{(2\theta_{12})}\right)+O(\varepsilon\times V_{matt})
\end{equation}
hence, the standard survival probability differs from the NSI-modified one by only a few percent.
}

\end{document}